\documentclass[iop]{emulateapj} %% the cool one!

\usepackage{graphicx}
\usepackage{natbib}
\usepackage{xcolor}

\def\lsim{\mathrel{\hbox{\rlap{\lower.55ex \hbox {$\sim$}}\kern-.0em
\raise.4ex \hbox{$<$}}}} 
\def\gsim{\mathrel{\hbox{\rlap{\lower.55ex \hbox {$\sim$}}\kern-.0em
\raise.4ex \hbox{$>$}}}}

\def\l{$\lambda$}

\def\c{$^{\rm c}$}

\shorttitle{Spectral evolution of AT~2018dyb}
\shortauthors{Leloudas et al.}

\begin{document}

\title{The spectral evolution of AT~2018dyb and the presence of metal lines in tidal disruption events}

\author{
Giorgos Leloudas\altaffilmark{1}, 
Lixin Dai\altaffilmark{2,3},
Iair Arcavi\altaffilmark{4,5},
Paul M. Vreeswijk\altaffilmark{6},
Brenna Mockler\altaffilmark{7,3},
Rupak Roy\altaffilmark{8},
Daniele B. Malesani\altaffilmark{9,3,1},
Steve Schulze\altaffilmark{10},
Thomas Wevers\altaffilmark{11},
Morgan Fraser\altaffilmark{12},
Enrico Ramirez-Ruiz\altaffilmark{7,3},
Katie Auchettl\altaffilmark{3},
Jamison Burke\altaffilmark{13,14},
Giacomo Cannizzaro\altaffilmark{6,15},
Panos Charalampopoulos\altaffilmark{3,1},
Ting-Wan Chen\altaffilmark{16},
Aleksandar Cikota\altaffilmark{17},
Massimo Della Valle\altaffilmark{18,19},
Lluis Galbany\altaffilmark{20}, 
Mariusz Gromadzki\altaffilmark{21},
Kasper E. Heintz\altaffilmark{22},
Daichi Hiramatsu\altaffilmark{13,14},
Peter G. Jonker\altaffilmark{6,15},
Zuzanna Kostrzewa-Rutkowska\altaffilmark{6,15},
Kate Maguire\altaffilmark{23,24},
Ilya Mandel\altaffilmark{25,26},
Matt Nicholl\altaffilmark{26,27},
Francesca Onori\altaffilmark{28},
Nathaniel Roth\altaffilmark{29},
Stephen J. Smartt\altaffilmark{23},
Lukasz Wyrzykowski\altaffilmark{21},
Dave R. Young\altaffilmark{23}
}

\altaffiltext{1}{DTU Space, National Space Institute, Technical University of Denmark, Elektrovej 327, 2800 Kgs. Lyngby, Denmark}
\altaffiltext{2}{Department of Physics, The University of Hong Kong, Pokfulam Road, Hong Kong}
\altaffiltext{3}{Dark Cosmology Centre, Niels Bohr Institute, University of Copenhagen, Lyngbyvej 2, 2100 Copenhagen, Denmark}
\altaffiltext{4}{The School of Physics and Astronomy, Tel Aviv University, Tel Aviv 69978, Israel}
\altaffiltext{5}{CIFAR Azrieli Global Scholars program, CIFAR, Toronto, Canada}
\altaffiltext{6}{Department of Astrophysics/IMAPP, Radboud University Nijmegen, P.O. Box 9010, 6500 GL Nijmegen, The Netherlands}
\altaffiltext{7}{Department of Astronomy and Astrophysics, University of California, Santa Cruz, CA 95064, USA}
\altaffiltext{8}{The Inter-University Centre for Astronomy and Astrophysics, Ganeshkhind, Pune - 411007, India}
\altaffiltext{9}{Cosmic DAWN centre,  Niels Bohr Institute, University of Copenhagen,  Lyngbyvej 2, 2100 Copenhagen, Denmark}
\altaffiltext{10}{Department of Particle Physics and Astrophysics, Weizmann Institute of Science, Rehovot 7610001, Israel}
\altaffiltext{11}{Institute of Astronomy, Madingley Road, Cambridge CB3 0HA, United Kingdom}
\altaffiltext{12}{School of Physics, O’Brien Centre for Science North, University College Dublin, Belfield, Dublin 4, Ireland}
\altaffiltext{13}{Department of Physics, University of California, Santa Barbara, CA 93106-9530, USA}
\altaffiltext{14}{Las Cumbres Observatory, 6740 Cortona Dr Ste 102, Goleta, CA 93117-5575, USA}
\altaffiltext{15}{SRON, Netherlands Institute for Space Research, Sorbonnelaan 2, 3584 CA, Utrecht, The Netherlands}
\altaffiltext{16}{Max-Planck-Institut f\"{u}r extraterrestrische Physik, Giessenbachstra\ss e, 85748 Garching, Germany}
\altaffiltext{17}{Physics Division, Lawrence Berkeley National Laboratory, 1 Cyclotron Road, Berkeley, CA 94720, USA}
\altaffiltext{18}{Ist. Nazionale di Astrofisica, Osservatorio Astronomico di Capodimonte (OACN), 80131 Napoli, Italy}
\altaffiltext{19}{European Southern Observatory, Karl-Schwarzschild-Strasse 2, D-85748 Garching, Germany}
\altaffiltext{20}{PITT PACC, Department of Physics and Astronomy, University of Pittsburgh, Pittsburgh, PA 15260, USA}
\altaffiltext{21}{Astronomical Observatory, University of Warsaw, Al. Ujazdowskie 4, 00-478 Warszawa, Poland}
\altaffiltext{22}{Centre for Astrophysics and Cosmology, Science Institute, University of Iceland, Dunhagi 5, 107, Reykjavik, Iceland}
\altaffiltext{23}{Astrophysics Research Centre, School of Mathematics and Physics, Queen's University Belfast, Belfast BT7 1NN, UK}
\altaffiltext{24}{School of Physics, Trinity College Dublin, Dublin 2, Ireland}
\altaffiltext{25}{Monash Centre for Astrophysics, School of Physics and Astronomy, Monash University, Clayton, Victoria 3800, Australia}
\altaffiltext{26}{Birmingham Institute for Gravitational Wave Astronomy, University of Birmingham, Birmingham, B15 2TT, UK}
\altaffiltext{27}{Institute for Astronomy, University of Edinburgh, Royal Observatory, Blackford Hill EH9 3HJ, UK}
\altaffiltext{28}{Istituto di Astrofisica e Planetologia Spaziali (INAF), Via Fosso del Cavaliere 100, Roma, I-00133, Italy}
\altaffiltext{29}{Department of Astronomy and Joint Space-Science Institute, University of Maryland, College Park, MD 20742, USA}

%%%%%%%%%%%%%%%%%%%%%%%%%%%%%%%%%%  ABSTRACT   %%%%%%%%%%%%%%%%%%%%%%%%%%%%%%

\begin{abstract}

We present light curves and spectra of the tidal disruption event (TDE) ASASSN-18pg / AT~2018dyb spanning a period of one year.
The event shows a plethora of strong emission lines, including the Balmer series, \ion{He}{2}, \ion{He}{1} and metal lines of 
\ion{O}{3} $\lambda$3760 and 
\ion{N}{3}  $\lambda\lambda$ 4100, 4640 (blended with \ion{He}{2}).
The latter lines are consistent with originating from the Bowen fluorescence mechanism.  
By analyzing literature spectra of past events, we conclude that these lines are common in TDEs.
The spectral diversity of optical TDEs is thus larger than previously thought and includes N-rich events besides H- and He-rich events. 
We study how the spectral lines evolve with time, by means of their width, relative strength, and velocity offsets. 
The velocity width of the lines starts at $\sim$ 13\,000 km s$^{-1}$ and decreases with time. 
The ratio of \ion{He}{2} to \ion{N}{3} increases with time. The same is true for ASASSN-14li, which has a very similar spectrum to AT~2018dyb but its lines are narrower by a factor of $>$2.
We estimate a black hole mass of 
$M_{\rm BH}$ = $3.3^{+5.0}_{-2.0}\times 10^6$ $M_{\odot}$
by using the $M$--$\sigma$ relation.
This is consistent with the black hole mass derived using the {\tt MOSFiT} transient fitting code.
The detection of strong Bowen lines in the optical spectrum is an indirect proof for extreme ultraviolet and (reprocessed) X-ray radiation and favors an accretion origin for the TDE optical luminosity. 
A model where photons escape after multiple scatterings through a super-Eddington thick disk and its optically thick wind, viewed at an angle close to the disk plane, is consistent with the observations. 

\end{abstract}




%%%%%%%%%%%%%%%%%%%%%%%%%%%%%%%%%%  INTRODUCTION   %%%%%%%%%%%%%%%%%%%%%%%%%%%%%%

\section{Introduction}

The optical spectra of tidal disruption events (TDEs) \citep{Rees88} are usually assumed to be dominated by H and He lines \citep{Arcavi14}.
The large diversity in the ratio of H to He  has been the topic of much discussion 
and has deep physical implications both for the nature of the disrupted star and for the radiative processes taking place during the event  
\citep{gezari12,Guillochon14,Roth16,Roth18}. 

On the other hand, metal lines are prominent in the UV spectra of TDEs \citep{Cenko16,Brown18}, with the most notable being those of highly ionized N, pointing to a possible relation to N-rich quasars \citep{kochanek16, Liu18}.
This raises the question of why such metal lines have not been unambiguously identified and reported in the optical regime. 
Indeed, \cite{Brown18} speculate
that, based on the UV spectra, the lines on the blue
side of \ion{He}{2} could be due to \ion{N}{3} and \ion{C}{3} similar to
features seen in Wolf--Rayet stars.
Recently, \cite{Blago15af} reported the detection of \ion{O}{3} and \ion{N}{3} lines in the spectrum of the TDE iPTF15af and attributed them to the mechanism of Bowen fluorescence. 

The Bowen fluorescence mechanism \citep{Bowen34, Bowen35} has been proposed  and widely discussed to explain the prominent \ion{O}{3} and \ion{N}{3} optical emission lines observed in various and diverse astrophysical systems such as planetary nebulae \citep[e.g.,][]{Unno55, Weymann69}, X-ray binary stars \citep[e.g.,][]{McClintock75}, and Wolf--Rayet stars \citep[e.g.,][]{Crowther07}.
It has also been proposed in the context of supermassive black hole accretion disks by \cite{1985ApJ...299..752N}, but only recently identified in a flaring AGN for the first time \citep{2019NatAs.tmp..187T}.
Since elements heavier than H and He should be rare in these systems, the only plausible explanation for seeing strong \ion{O}{3} and \ion{N}{3} fluorescent lines is that these lines are excited by some large sources of energy not available to the predominant H and He. In this mechanism, the eventual \ion{O}{3} and \ion{N}{3} optical emission features result from a series of processes. The first process is the ionization of singly ionized He (\ion{He}{2}) by photons with wavelength shorter than 228 \AA. During the recombination process of fully ionized \ion{He}{2}, while the transition of one electron between the outer orbits can give rise to the optical lines such as \ion{He}{2} at $\lambda$4686, the final transition from $n=2$ to $n=1$ (\ion{He}{2}~Ly$\alpha$) produces an extreme ultraviolet (EUV) line at 304~\AA. A secondary process is the excitation of a few \ion{O}{3} and \ion{N}{3} states by the absorption of the intense EUV photons at 304 \AA, because these ions have a transition very close to this wavelength by coincidence. These excited ions then return to the ground state through a series of transitions, producing optical lines such as \ion{O}{3} at $\lambda\lambda$ 3047, 3133, 3312, 3341, 3444, 3760 and \ion{N}{3} at $\lambda\lambda$ 4097, 4104, 4379, 4634, 4641 \citep{Osterbrock74}.

In this paper we present observations of the TDE ASASSN-18pg / AT~2018dyb which shows strong Bowen fluorescent N and O lines.  
At the same time, we revisit the spectra of past TDEs and we show that similar lines are conspicuous in many of them. Our data are presented in 
Section~\ref{sec:obs}. The spectral evolution is described in Section~\ref{sec:res} and we focus on the analysis of the emission lines in Section~\ref{sec:emlines}. Section~\ref{sec:highres} focuses on early-time high-resolution spectroscopy 
and Section~\ref{sec:medres} on late-time medium-resolution spectroscopy and the determination of the black hole mass. 
In Section~\ref{sec:host} we study the host galaxy and discuss host contamination, while in Section~\ref{sec:mosfit} we model the light curves to extract fundamental properties of the TDE and its progenitor system. Section~\ref{sec:disc} contains our discussion and Section~\ref{sec:conc} summarizes our conclusions.

%%%%%%%%%%%%%%%%%%%%%%%%%%%%%%%%%%  OBSERVATIONS   %%%%%%%%%%%%%%%%%%%%%%%%%%%%%%

\section{Observations} \label{sec:obs}

ASASSN-18pg / AT~2018dyb was discovered by the All-Sky Automated Survey for Supernovae  \citep[ASAS-SN;][]{ASASSN} 
on 2018-07-11 \footnote{all dates are in UT.} (first detection at $V = 16.5$ mag), with the last non-detection ($V>17.5$ mag) being nine days earlier \citep[][]{TNSdisc}. 
It was classified as a TDE  by \cite{TNSclas} on 2018-07-17, based on a spectrum obtained at the SOAR telescope. 
Our own observations set the precise redshift to $z = 0.0180$ (see Section~\ref{sec:highres}), which is used throughout this paper.
From now on we will primarily refer to AT~2018dyb with its TNS name \footnote{The TNS is the official IAU mechanism for reporting new astronomical transients: \url{https://wis-tns.weizmann.ac.il/}}.

\subsection{Archival host galaxy observations and constraints on the nuclear nature of AT~2018dyb}

The location of AT~2018dyb was observed by the SkyMapper survey in 2015 \citep{SkyMapper}. 
We retrieved the SkyMapper images, \footnote{\url{http://skymapper.anu.edu.au/}} and the host galaxy is detected in the $ugriz$ filters (only marginally in $u$). It is also detected in the Two Micron All Sky Survey (2MASS) survey, although blended with a nearby star of similar brightness. 

The archival host detection allows  constraints to be placed on the offset between host and transient. To measure the position of AT~2018dyb, we used a UVOT image in the UVW2 filter, taken close to maximum light (see Sect.~\ref{sec:UVOT}), because there is negligible host contribution in the UV bands around that epoch.
We measure $\mathrm{RA} = 16^{\rm h} 10^{\rm m} 58\fs86$, $\mathrm{Dec} = -60\degr 55\arcmin 24\farcs28$ for the transient, calibrated against the \textit{Gaia} survey catalog (we note that our position is $\approx 1.3\arcsec$ away from the one reported by \citealt{TNSdisc}). To accurately compute the offset, we cross-registered the UVOT and the SkyMapper images, which results in an RMS scatter of $\approx 0.25\arcsec$ in both RA and Dec, based on 26 common sources. This rather large scatter stems from the fairly wide point-spread function (PSF) of the SkyMapper images ($\approx 2.5-3\arcsec$). 
For the host galaxy we measure $\mathrm{RA} = 16^{\rm h} 10^{\rm m} 58\fs89$, $\mathrm{Dec} = -60\degr 55\arcmin 24\farcs46$, i.e. an offset of $0\farcs21$ in RA and $0\farcs18$ in Dec. The transient is thus consistent with the location of the nucleus within our measurement uncertainties. 
For reference, $0.25\arcsec$ corresponds to 100~pc at $z = 0.0180$.
 
\subsection{UVOT photometry}
\label{sec:UVOT}

\begin{figure}
\includegraphics[width=\columnwidth]{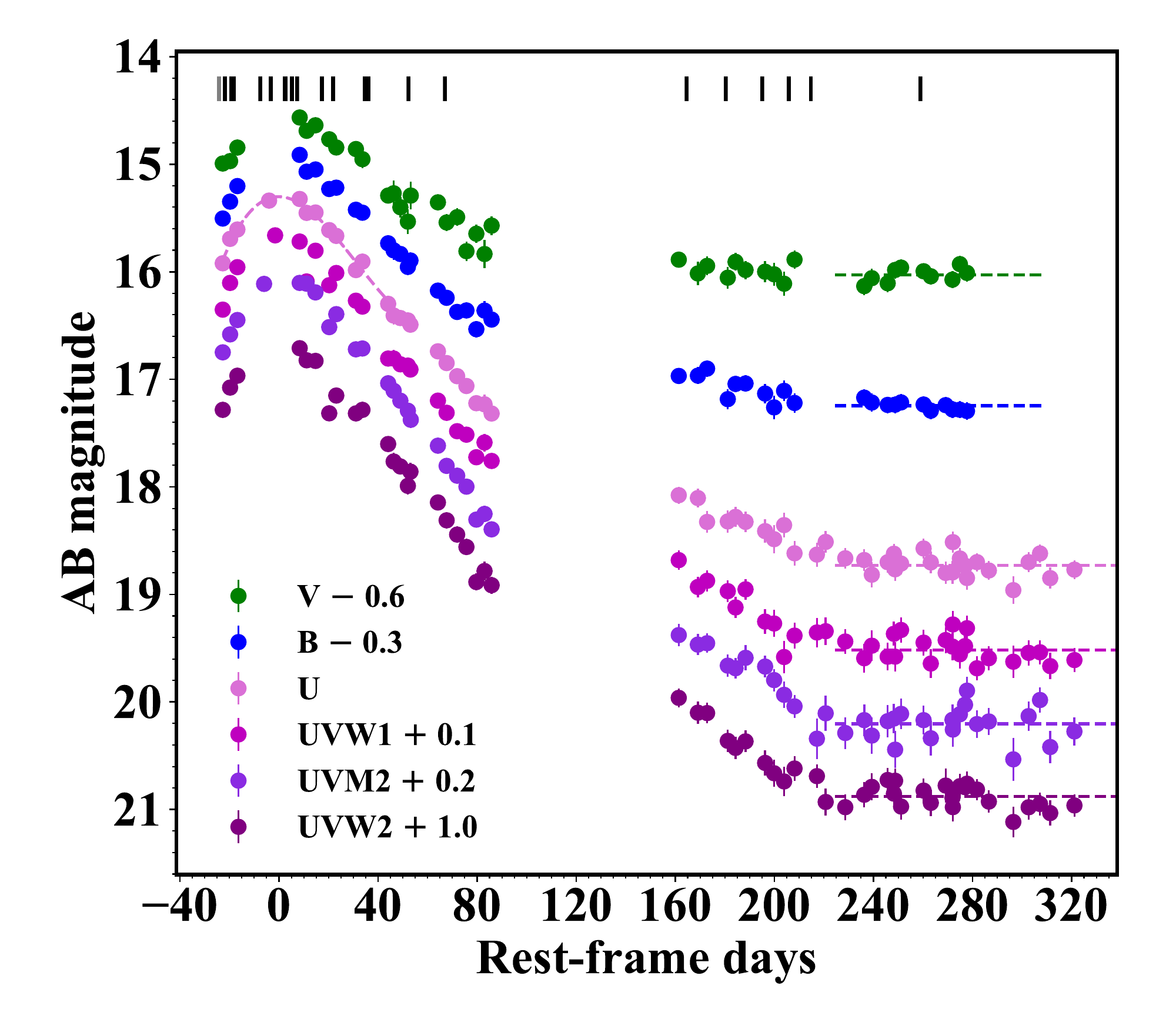}
\caption{The UVOT light curves of AT~2018dyb shown in rest-frame days with respect to the date of maximum light  in the $U$ band ($\mathrm{MJD} = 58340.74$), as obtained by a polynomial fit. The light curves are shifted for clarity as indicated in the legend. Vertical bars denote epochs of spectroscopic observations. Horizontal dashed lines indicate the host level as computed by the flattening of the light curves after $+230$ days.
\label{fig:UVOT}}
\end{figure}

Photometry of AT~2018dyb was obtained by the Neil Gehrels \textit{Swift} Observatory between 2018-07-18 and 2019-07-03 in 60 epochs. The transient was bright in all near-UV (NUV) and optical filters of the \textit{Swift}/UVOT telescope. The UVOT data were reduced using the standard pipeline available in the HEAsoft 
software package \footnote{\url{https://heasarc.nasa.gov/lheasoft/}}. 
Observation of every epoch was conducted using one or several orbits. To improve the signal-to-noise ratio of the observation in a given band in a particular epoch, we co-added all orbit-data for that corresponding epoch using the HEAsoft routine \texttt{uvotimsum}. We used the routine \texttt{uvotdetect} to determine the correct position of the transient (which is consistent with the ground-based optical observations) and used the routine \texttt{uvotsource} to measure the apparent magnitude of the transient by performing aperture photometry. Late time images reveal the presence of multiple contaminating sources near the transient and its host, so for source extraction we used a small aperture of radius 3$\arcsec$, while an aperture of radius 100$\arcsec$ was used to determine the background. Our photometry is listed in Table~\ref{tab:UVOTphot} and on the AB system. The light curves of AT~2018dyb are shown in Figure~\ref{fig:UVOT}. Maximum light in the $U$ band occurred at $\mathrm{MJD} = 58340.74$ (Figure~\ref{fig:UVOT}). 
This is the reference date that is adopted for all phases quoted throughout the paper.

\subsection{Las Cumbres and ePESSTO spectroscopy}

\begin{figure*}
\includegraphics[width=\textwidth]{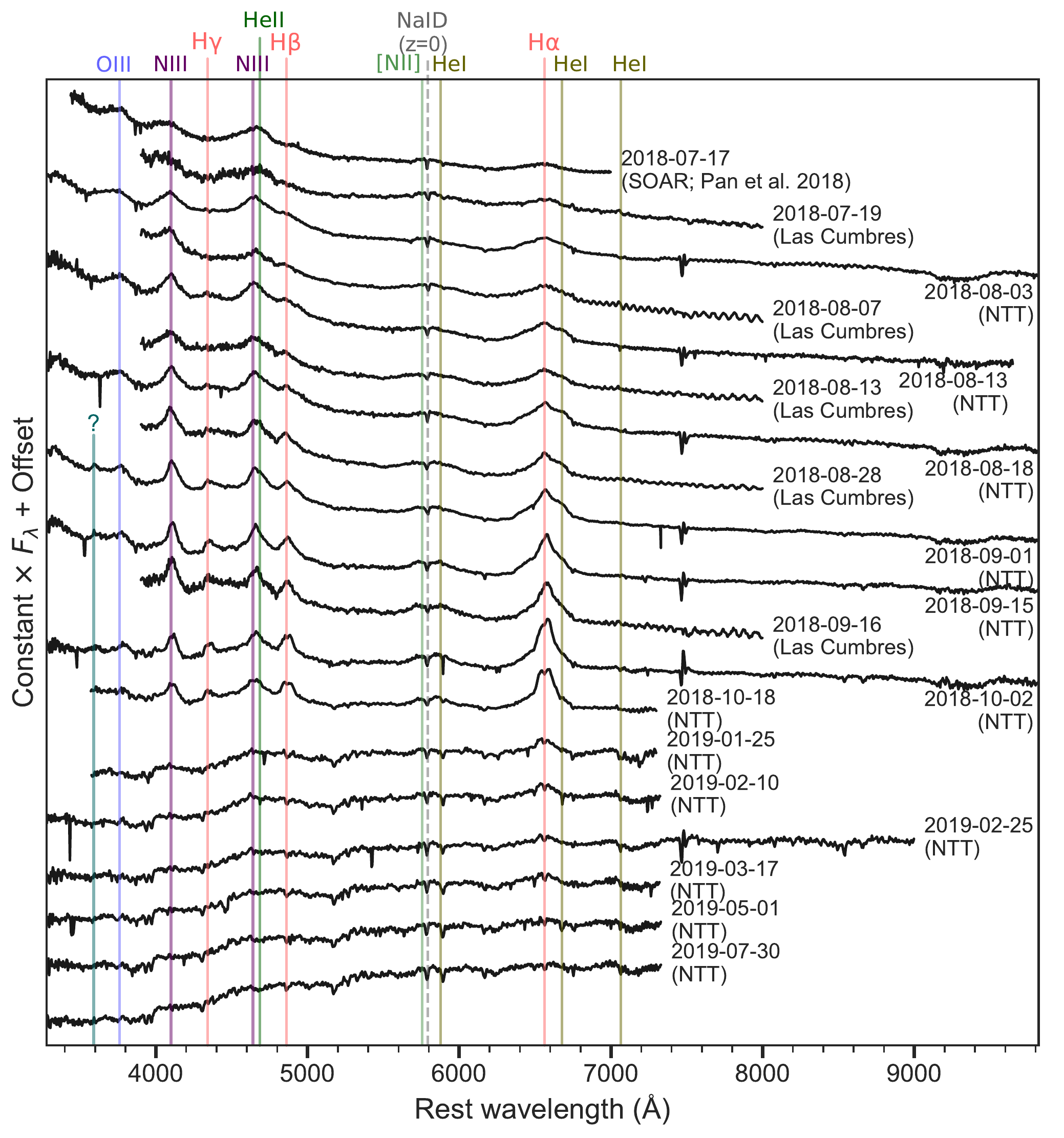}
\caption{The low-resolution spectra of AT~2018dyb with the strongest broad lines identified. The SOAR spectrum \citep{TNSclas} has been retrieved from the TNS. 
\label{fig:specseq}}
\end{figure*}

We have collected low-resolution spectra of AT~2018dyb using the FLOYDS instrument on the Las Cumbres 
Observatory \footnote{\url{http://lco.global}} (LCO) 2 m telescope in Siding Spring,
Australia \citep{LCO} and with EFOSC2 on the New Technology Telescope (NTT) in La Silla Observatory, Chile, as part of the ePESSTO survey \citep{pessto}. 
The LCO spectra were reduced using the pyraf-based \texttt{floydsspec} pipeline originally developed by S. Valenti. The NTT spectra were reduced in a standard manner with the aid of the PESSTO pipeline \citep{pessto}. A spectroscopic log can be found in Table~\ref{tab:speclog} and the spectral series is shown in Figure~\ref{fig:specseq}. 
All spectra are available through the WISeREP archive \citep{WISeREP}.

\subsection{UVES spectroscopy}

In addition, we obtained high-resolution spectroscopy with the Ultraviolet and Visual Echelle Spectrograph (UVES) mounted on the Kueyen unit of ESO's Very Large Telescope (VLT).
AT~2018dyb was observed on the nights of 2018-07-22, 2018-07-23 and 2018-08-16.
The dichroic/central wavelength settings 346+580 nm (dichroic 1) and
437+860 nm (dichroic 2) were used, covering the full optical range 
305-1040~nm, with the exception of small gaps at 575.5-583.8~nm and
851.7-867.4~nm, in two exposures. During the night of 2018-07-22, only a
single setting (346+580 nm) could be obtained through thick cirrus before the telescope had to be closed. 
The data were reduced with the UVES pipeline \citep{Ballester00} and the spectral trace, wavelength calibration solution, and final extraction were inspected and slightly improved by adjusting some of the pipeline recipe parameters. 
The rms of the wavelength solution varied between 0.001 and 0.005~\AA. 
The incomplete data from 2018-07-22 were co-added with  the data from 2018-07-23 to produce a single spectrum covering the whole wavelength range.

\subsection{X-shooter spectroscopy}

On 2019-03-08, 206 days after the $U$-band maximum, we obtained a spectrum of AT~2018dyb with the medium-resolution spectrograph X-shooter on the VLT \citep{2011A&A...536A.105V}. 
We used the nodding along the slit mode, completing one `ABBA' cycle. 
The spectrum was reduced with the dedicated EsoReflex pipeline (v.~3.3.4). 
The resulting spectrum covers the combined wavelength range 3200 -- 24700~\AA\ with a nominal resolving power of $R = 5400$ in the UVB and $8900$ in the VIS arm.  

\subsection{X-Rays}

We finally note that AT~2018dyb does not exhibit strong X-ray emission.
Using the \textit{Swift} XRT observation at time of discovery, \cite{2018ATel11876....1M} derived a conservative upper limit to the absorbed flux of
$7 \times 10^{-14}$~erg~cm$^{-2}$~s$^{-1}$
(0.003 counts~s$^{-1}$) in the 0.3--10.0~keV band, assuming a blackbody with a $kT=0.2$~keV. 
By merging the first 14 epochs of the Swift XRT observations, we derive a slightly lower upper limit of $1.6 \times 10^{-3}$~counts~s$^{-1}$. 
Using a blackbody model with $kT=0.2$ keV and a Galactic column density of $N_\mathrm{H} \sim 1.83 \times 10^{21}$~cm$^{-2}$, we derive an upper limit for the absorbed flux of 
$3.4 \times 10^{-14}$~erg~cm$^{-2}$~s$^{-1}$.
This corresponds to an unabsorbed flux of $7.8 \times 10^{-14}$~erg~cm$^{-2}$~s$^{-1}$ and, at the redshift of the host, to a luminosity of $\sim 5.6 \times 10^{40}$~erg~s$^{-1}$, which suggests no evidence of a strong active galactic nucleus \citep[e.g.,][]{2006A&A...451..457T, 2017ApJS..232....8L,2017ApJS..233...17R}.

%%%%%%%%%%%%%%%%%%%%%%%%%%%%%%%%%%  RESULTS  SPECTROSCOPY  %%%%%%%%%%%%%%%%%%%%%%%%%%%%%%

\begin{figure*}
\includegraphics[width=\textwidth]{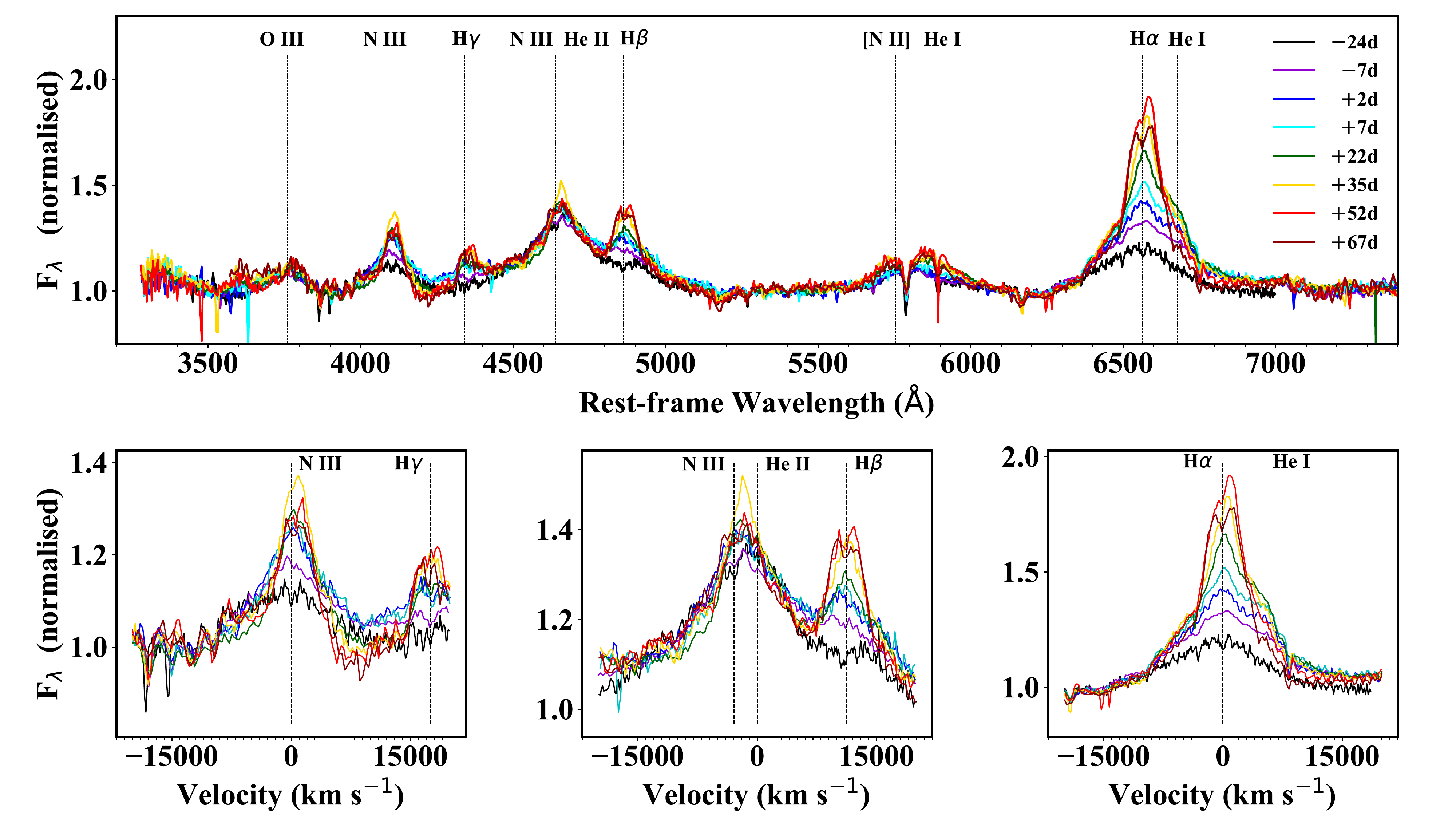}
\caption{\textbf{Upper panel:} selected spectra of AT~2018dyb after normalising with the continuum, overplotted with different colors as indicated in the legend. The phases refer to the time of $U$-band maximum and they are given in the rest frame. The strongest lines have been identified. 
All lines grow stronger with time and this is particularly true for the Balmer lines. 
\textbf{Lower panels:} a zoom-in on some of the strongest lines in velocity space. In the middle panel, the zero velocity has been set considering \ion{He}{2}, but it can be seen that the line peaks between \ion{He}{2} and \ion{N}{3} and that it is likely a blend. The width of the lines starts at  $\sim$12,000 km~s$^{-1}$ and decreases with time (see also Figure~\ref{fig:lineevol2018dyb}).
\label{fig:specnorm}}
\end{figure*}

\section{Spectral evolution} \label{sec:res}

The first spectra of AT~2018dyb are dominated by H$\alpha$ and three more broad features centered at 4660, 4100, and 3760 \AA.
At these phases, H$\beta$ is weak and H$\gamma$ is not detected, providing the first evidence that the line at 4100 \AA{} is unlikely to be H$\delta$ as previously identified in TDEs with similar spectra \citep[e.g.][]{Holoien14li,Hung16axa}.
As time passes, the spectrum becomes redder and the lines become narrower. At the same time, H$\beta$ and H$\gamma$ emerge and become stronger with time, but remain weaker than the 4100 \AA{} line.
We propose here that this strong line is dominated by \ion{N}{3} $\lambda$4100  (blend of $\lambda\lambda$4097, 4104).
In addition, we identify the bluer broad feature as \ion{O}{3} $\lambda\lambda$ 3754, 3757, and 3759, a blend of the strongest \ion{O}{3} lines in the optical \citep[National Institute of Standards and Technology, Atomic Spectra Database Lines Data;][]{nist}, which from now on we will call \ion{O}{3} $\lambda$3760 for simplicity. 
The simultaneous detection of these \ion{N}{3} and \ion{O}{3} lines is compatible with the idea that they might originate from Bowen fluorescence \citep{Blago15af,2019NatAs.tmp..187T} and strengthens these identifications (because these lines are expected to appear together due to the same physical mechanism).
In addition, this identification suggests that the broad line at 4660 \AA{} is (at least partially) due to \ion{N}{3} $\lambda$4640 and not (exclusively) \ion{He}{2} $\lambda$4686, as usually assumed.

Together with the emergence of H$\beta$ and H$\gamma$ we have the appearance of a shoulder in the red wing of H$\alpha$, which can be attributed to \ion{He}{1} $\lambda6678$. 
This raises the question of whether the last strong feature apparent in the optical spectrum, between 5690 and 5890~\AA, can be due to \ion{He}{1} $\lambda5876$. 
It is unlikely that this is the sole contribution to this feature, given that \ion{He}{1} would need to be blueshifted by 5000 km s$^{-1}$, while the other lines in the spectrum appear close to rest velocity. 
\cite{Blago15af} suggested an alternative identification for this line as [\ion{N}{2}] $\lambda5754$, but this association is not secure either. 
We stress that the profile of this line is complicated and it is strongly affected by strong Galactic \ion{Na}{1}~D absorption. So we tentatively identify this feature as a blend of the above lines. 
Finally, one feature remains unidentified -- a weak line at 3590 \AA{}, which appeared after 2019-09-01 to the blue of the \ion{O}{3} blend.

AT~2018dyb disappeared behind the Sun around mid-October 2018. The first spectrum after it re-appeared in 2019 January shows that most features have weakened or disappeared in the meantime. The spectrum resembles mostly that of an elliptical galaxy with a superimposed broad H$\alpha$ component, due to the TDE, and some residual flux in the \ion{N}{3}/\ion{He}{2} region, although individual broad features can no longer be clearly identified there. At the same time, the TDE has faded by more than 1~mag in the UV bands.
The later spectra become even redder, while H$\alpha$ progressively disappears.

\section{Emission Line Analysis} \label{sec:emlines}

In order to focus on the emission lines, we have first de-reddened and then removed the continuum from the spectra.
The spectra were de-reddened for the Galactic extinction in the direction of AT~2018dyb of $A_V = 0.625$~mag \citep{2011ApJ...737..103S}, while no extinction was assumed for the (passive) host galaxy. These choices are discussed further below. 
Subsequently, we fit a low-order polynomial in the regions of the spectra that we estimated were continuum-dominated. While this procedure is subjective to some degree, it is standard practice in the literature. 
The AT~2018dyb spectra have not been host subtracted, so we stress that this procedure removes the total continuum (TDE+host) and could leave some line contamination from the host.
However, inspecting the last spectra obtained after January, we estimate that this line contamination is small. This analysis was only attempted in the `early' spectra (up to 70 days past maximum and before the TDE disappeared behind the Sun in mid-October).

By dividing by the continuum we obtain the normalised spectra that can be seen in Figure~\ref{fig:specnorm}. 
In this figure it is immediately apparent that the emission lines grow stronger with time with respect to the continuum (their equivalent width gets larger). The lower panels zoom on the strongest features and show their evolution in velocity space. The FWHM of the lines is approximately $\sim$13,000 km~s$^{-1}$ at discovery and decreases with time. In the last two spectra, an absorption component becomes visible in the middle of the H$\alpha$ and H$\beta$ lines (at $v = 0$ km~s$^{-1}$). This is a feature of the host galaxy (stellar absorption) that becomes relatively more important as the host galaxy contribution increases with time. However, these features are weak compared to other TDEs where they dominate the spectrum even at maximum light 
\cite[e.g.][]{Holoien14li,Blago16fnl}. This is due to the fact that the host of AT~2018dyb is not an E+A galaxy as has been observed for many TDEs \citep{Arcavi14,2016ApJ...818L..21F}.

\begin{figure}
\includegraphics[width=\columnwidth]{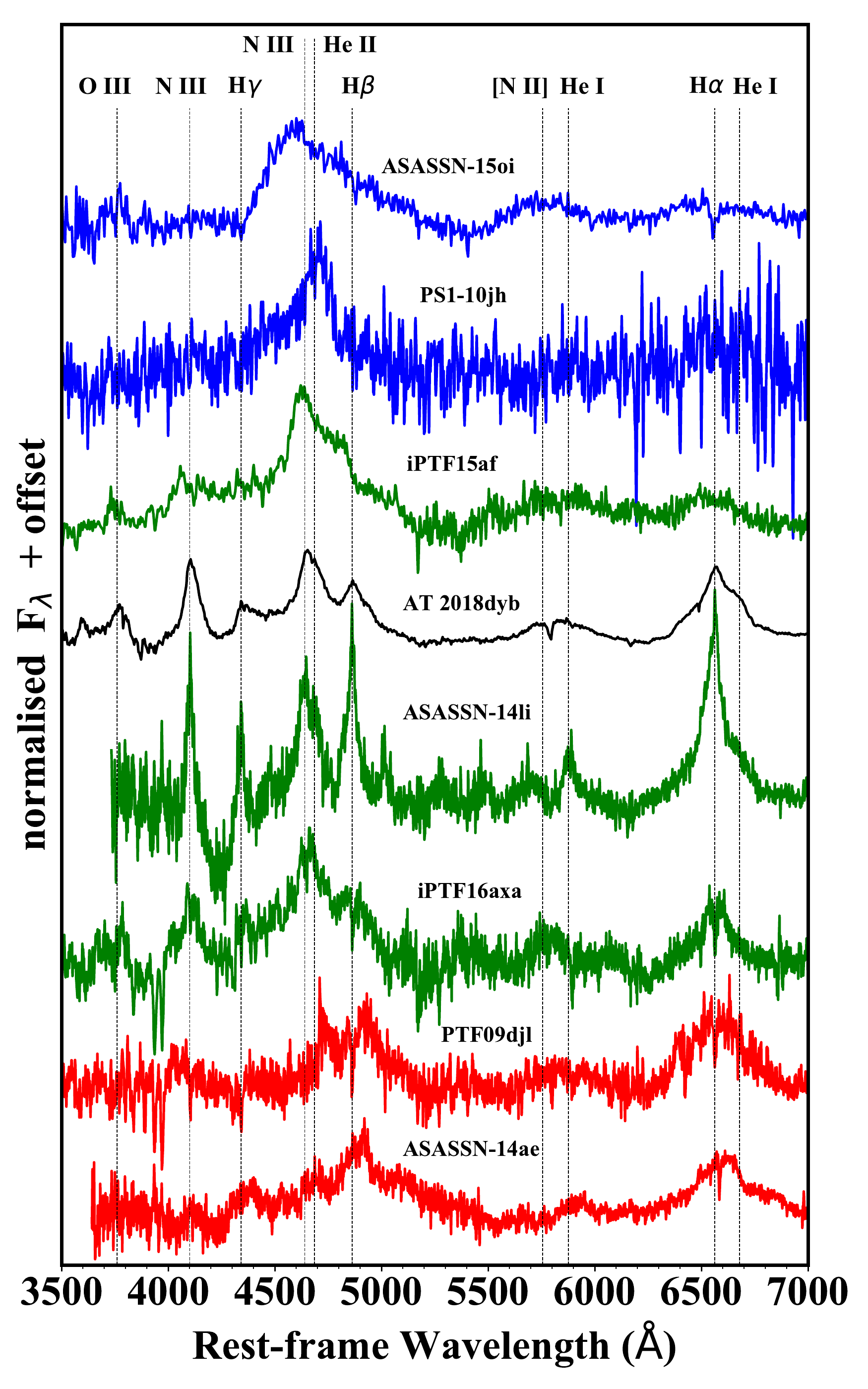}
\caption{AT~2018dyb compared to other TDEs.  
This is a version of the original plot by \cite{Arcavi14}, where we focus on the existence of events with strong N and O lines (`N-rich').
The top spectra belong to two `He-rich' events that do not show any evidence for H lines \citep{Holoien15oi,gezari12}. 
The next four spectra show TDEs that, in addition to \ion{He}{2}, show a clear peak at 4640 \AA\ and/or a very strong line at 4100 \AA\ (both attributed to \ion{N}{3}) and/or another emission line at $\sim$3760 \AA\ (\ion{O}{3}).
The spectra of AT2018dyb and ASASSN-14li are very similar, but the lines of ASASSN-14li are narrower. 
Finally, the bottom two spectra  \citep{Arcavi14,Holoien14ae} show weak or absent \ion{He}{2} (or N and O) lines, especially in comparison to their strong H lines. 
\label{fig:speccomp}}
\end{figure}

Figure \ref{fig:speccomp} shows a comparison of AT~2018dyb with other TDEs. 
The top spectra (blue) are `He-rich' with no evidence for H lines or other strong features. In addition, 
the broad \ion{He}{2} line of ASASSN-15oi \citep{Holoien15oi} appears clearly blueshifted, while for PS1-10jh this line appears almost at rest velocity, albeit with a visible blue wing \citep[as also noted by][]{gezari12}.
The bottom spectra (red) show weak or no \ion{He}{2} lines, while the dominant features are Balmer lines \citep{Arcavi14,Holoien14ae}. 
AT~2018dyb is more similar to the middle (green) spectra \citep{Holoien14li,Hung16axa} that show the following properties: a clear peak at 4640 \AA\ and/or a very strong line at 4100 \AA\ (both attributed to \ion{N}{3}) and/or another emission line at $\sim$3760 \AA\,, attributed to \ion{O}{3}. 
We observe that the spectrum of AT~2018dyb is very similar to ASASSN-14li with the main difference being that the lines are 
narrower in ASASSN-14li. In fact, for this TDE, \ion{N}{3} $\lambda4640$ and \ion{He}{2} $\lambda4686$ are clearly resolved into two peaks, confirming that (i) the identification of \ion{N}{3} is solid and (ii) both lines are approximately at zero velocity.
iPTF16axa also has the same set of N and O lines and so does 
iPTF15af \citep{Blago15af}, showing that these metal lines are common in TDEs. A subset of TDEs are therefore "N-rich".

\begin{figure}
\includegraphics[width=\columnwidth]{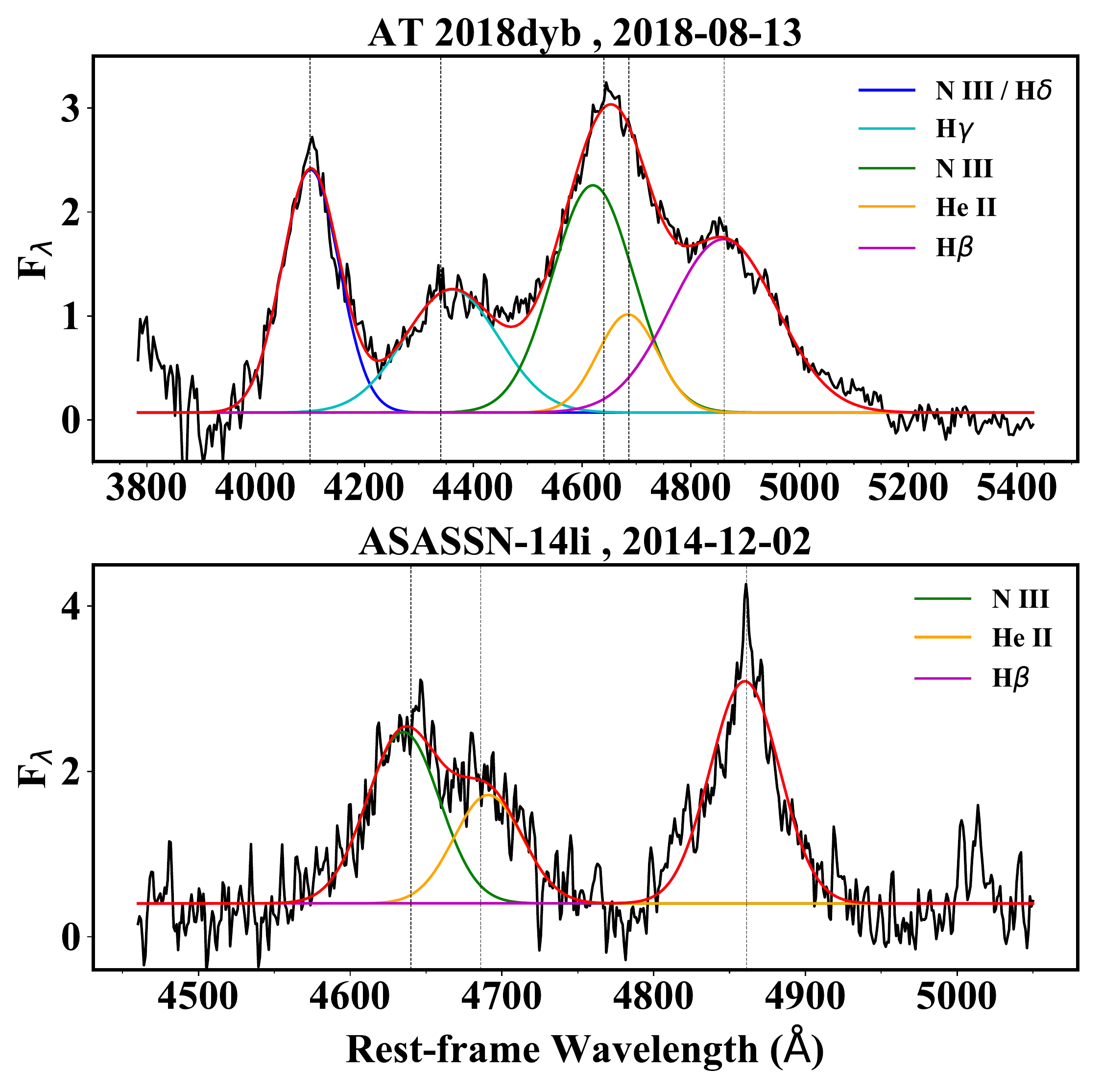}
\caption{\textbf{Upper panel:}
example of a simultaneous fit with five Gaussians in the spectrum of AT~2018dyb. The central wavelengths of the lines at zero velocity have been marked with dashed lines.
\textbf{Lower panel:}
similar but for ASASSN-14li. For this TDE the lines are narrower and do not blend as severely as for AT~2018dyb. For this reason, it not necessary to fit five lines simultaneously. Here, we show a triple fit focusing on \ion{N}{3}, \ion{He}{2} and H$\beta$.
\label{fig:exampleGfits}}
\end{figure}
 
\begin{figure*}
\includegraphics[width=\textwidth]{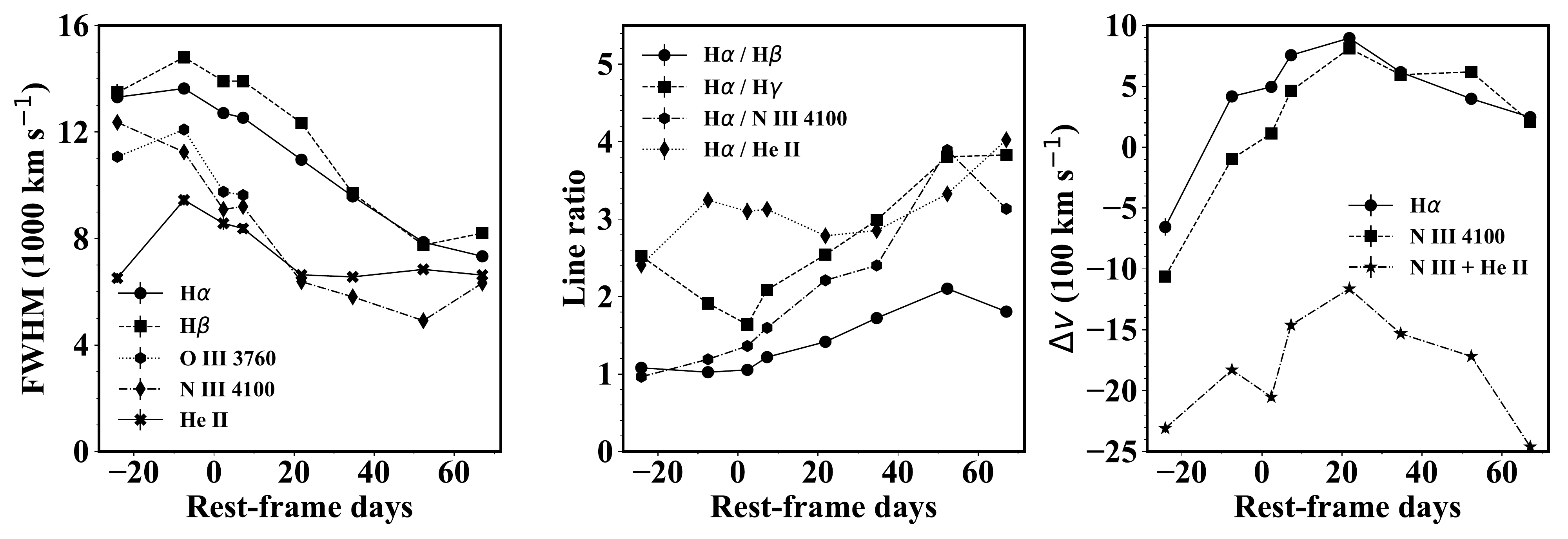}
\caption{Evolution of line widths (left), line ratios (middle), and line velocity offsets (right) for emission lines in the spectra of AT~2018dyb with respect to the date of $U$-band maximum. In the last panel, the fit to the \ion{N}{3} + \ion{He}{2} blend has been done with a single Gaussian and assuming the reference wavelength for \ion{He}{2}. This has been done to illustrate that, under these assumptions, this line would be significantly blueshifted with respect to the other strong lines, presenting additional evidence that this is a blend with  \ion{N}{3}.
\label{fig:lineevol2018dyb}}
\end{figure*}

\begin{figure*}
\includegraphics[width=\textwidth]{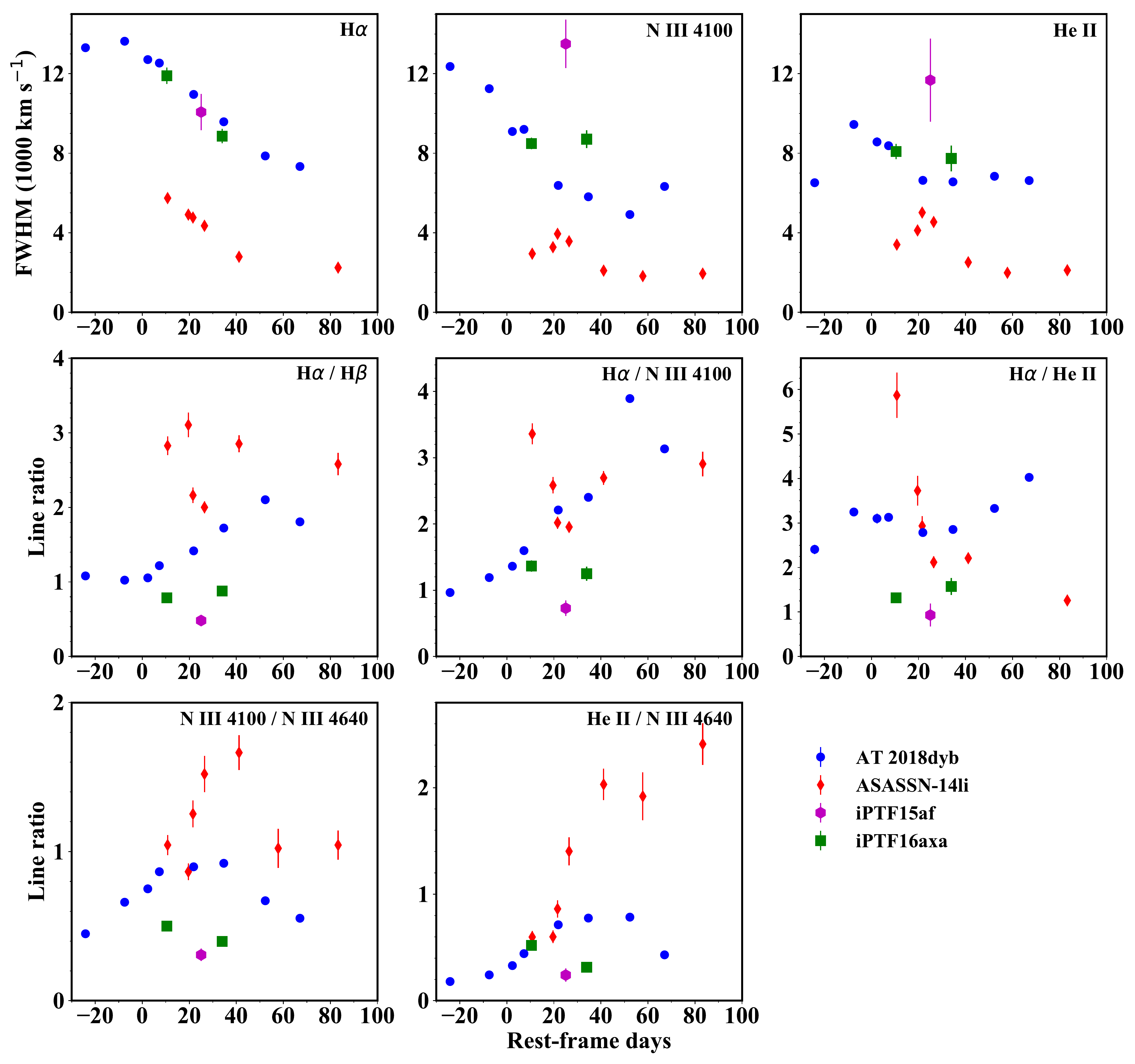}
\caption{Comparison of AT~2018dyb with other N-rich TDEs. The upper panels show the width evolution for different emission lines.
In the second and third rows, we show the evolution of selected line ratios. The symbols and colors used are explained in the legend  in the bottom right corner. The time is given in rest-frame days with respect to maximum light (AT~2018dyb and iPTF15af) or with respect to the time of discovery when the peak is unconstrained (ASASSN-14li and iPFT16axa). 
\label{fig:lineevolcomp}}
\end{figure*}

To get a quantitative view of the line evolution, we fit the emission lines with Gaussian profiles. This is done with custom-made routines based on \texttt{mpfit} \citep{mpfit}.
First, H$\alpha$ is fit simultaneously with \ion{He}{1} $\lambda$6678.
In the blue part of the spectrum, where many lines are blended together, we fit five lines simultaneously, namely the line at 4100 \AA, H$\gamma$, \ion{N}{3}, \ion{He}{2}, and  H$\beta$. 
This is a fit with many free parameters and we therefore impose some reasonable constraints in order to include some physical information and reduce the number of possible solutions. 
We require that H$\beta$ and H$\gamma$ have similar FWHM to H$\alpha$ (within 2000 km s$^{-1}$).
Similarly, we constrain the FWHM of the two \ion{N}{3} lines to be the same. 
We also allow only limited velocity shifts for the central wavelengths of the lines. 
An example fit for this complex region can be seen in Figure~\ref{fig:exampleGfits}.
It should be noted that the region around H$\gamma$ is particularly complex, and assuming that it can be modeled by a single Gaussian is probably an oversimplification. One possibility is that \ion{N}{3} $\lambda$4379 contributes significantly in this region, especially given the identification of other \ion{N}{3} Bowen lines in the spectrum. Nevertheless, for simplicity, we have decided not to include more lines in our fit. 
Tables~\ref{tab:linelum} and \ref{tab:FWHM} contain the line fit results.

The left panel of Figure~\ref{fig:lineevol2018dyb} shows the FWHM evolution of the emission lines in AT~2018dyb.  
We observe that the H$\alpha$ and the $\lambda$4100 line (assumed to be \ion{N}{3}) have a similar evolution, starting from a width of 13\,000 km~s$^{-1}$ 20 days before maximum and decreasing gradually to 6--8\,000 km~s$^{-1}$ almost 90 days later. 
The width of the \ion{O}{3} $\lambda$3760 line is similar at early times but the line profile evolves to become more complicated later and can no longer be fit by a single Gaussian. 
With the exception of the first epoch,  the width of \ion{He}{2} also shows a decreasing trend.
We should note that while the error bars we present include a proper propagation of the error spectrum and the errors in the line fits (resulting from \texttt{mpfit}), they do not include any estimate for the uncertainty during the procedure of removing the continuum. This systematic error is likely the most dominant and thus the errors in Figure~\ref{fig:lineevol2018dyb} are underestimated. 

The middle panel of Figure~\ref{fig:lineevol2018dyb} shows the evolution of several line ratios for AT~2018dyb. Absolute line luminosities are less reliable because they are more prone to uncertainties in the reddening correction.
We observe that the H$\alpha$/H$\beta$ ratio starts close to one. 
There is some evidence that this ratio is increasing with time, but it  always stays below the value expected in case B recombination for zero extinction. 
The ratio H$\alpha$/$\lambda$4100 also starts close to unity and increases with time until H$\alpha$ becomes four times stronger. 
For AT~2018dyb,  H$\alpha$ is about three times stronger than \ion{He}{2} but this is after removing the \ion{N}{3} contribution.

Finally, in the right panel of Figure~\ref{fig:lineevol2018dyb} we show the velocity shifts of some lines with respect to their rest-frame velocity. We focus only on the H$\alpha$ and the \ion{N}{3} $\lambda$4100, which are the most isolated lines and give the most reliable results. For many other lines the exact shifts depend on the details of the (quintuple) line fit and/or are part of the fit constraints themselves. However, these two lines always give consistent  results. It is very interesting that both lines start with a blueshift of 800-1\,000 km~s$^{-1}$ but quickly shift to the red (again by 800 km~s$^{-1}$) on a  timescale of 50 days. Subsequently, their line centre seems to return again toward rest velocity. 
Also included is a joint fit to the \ion{N}{3}+\ion{He}{2} blend with a single Gaussian, assuming that this is solely \ion{He}{2}  (i.e.,~assuming 4686 \AA\ as reference wavelength). This line would be significantly blueshifted with respect to the other lines by a consistent offset of 2000 km~s$^{-1}$. This presents additional proof that this line is a blend and not simply \ion{He}{2}.

For comparison, we fit the same set of lines in ASASSN-14li, iPTF15af, and iPTF16axa. 
In the case of ASASSN-14li, the lines are narrower and resolved and it is not necessary to make a simultaneous fit for all five lines in the blue (Figure~\ref{fig:exampleGfits}).
For iPTF15af and iPTF16axa we restrict ourselves to the highest quality spectra, because it was not possible to obtain reliable quintuple  fits for all spectra. We stress that deblending \ion{N}{3} and \ion{He}{2} is not trivial, and the result depends to a certain degree on the choices and constraints adopted for the simultaneous  fit. It was therefore important to fit these TDEs in a systematic way, consistently with AT~2018dyb (using the same choices as above), and therefore the fit results could be different than those presented in \cite{Blago15af} and \cite{Hung16axa}. The evolution of the different TDEs is compared to AT~2018dyb in Figure~\ref{fig:lineevolcomp}.

The line widths of ASASSN-14li show a similar decreasing behavior \citep{Holoien14li}  but they are substantially smaller than in AT~2018dyb, evolving from 6\,000 to 2\,000 km~s$^{-1}$. This is why \ion{N}{3} and \ion{He}{2} can easily be resolved in this TDE, while a deblending is needed for AT~2018dyb and the iPTF events. 
Interestingly, the widths of both \ion{N}{3} $\lambda$ 4100 and \ion{He}{2} in ASASSN-14li  show an initial increase (for the first 30 days), which is not the case for the Balmer lines. 
We note that the FWHM in the deblended \ion{He}{2} line of AT~2018dyb also shows a similar evolution. 
In terms of line widths, the iPTF events are more similar to AT~2018dyb than ASASSN-14li. 

The line ratios of H$\alpha$ to H$\beta$, \ion{N}{3}  and \ion{He}{2} show a large spread for the different TDEs. 
While ASASSN-14li has a value closer to three for H$\alpha$/H$\beta$, we observe that this is not a general rule and that TDEs can often have a ratio that is closer to one. The recent study of the fast-evolving iPTF16fnl, which was also found to be N-rich by \cite{Onori16fnl}, confirms the diversity in the observed line ratios and evolution.

\begin{figure*}
\includegraphics[width=\textwidth]{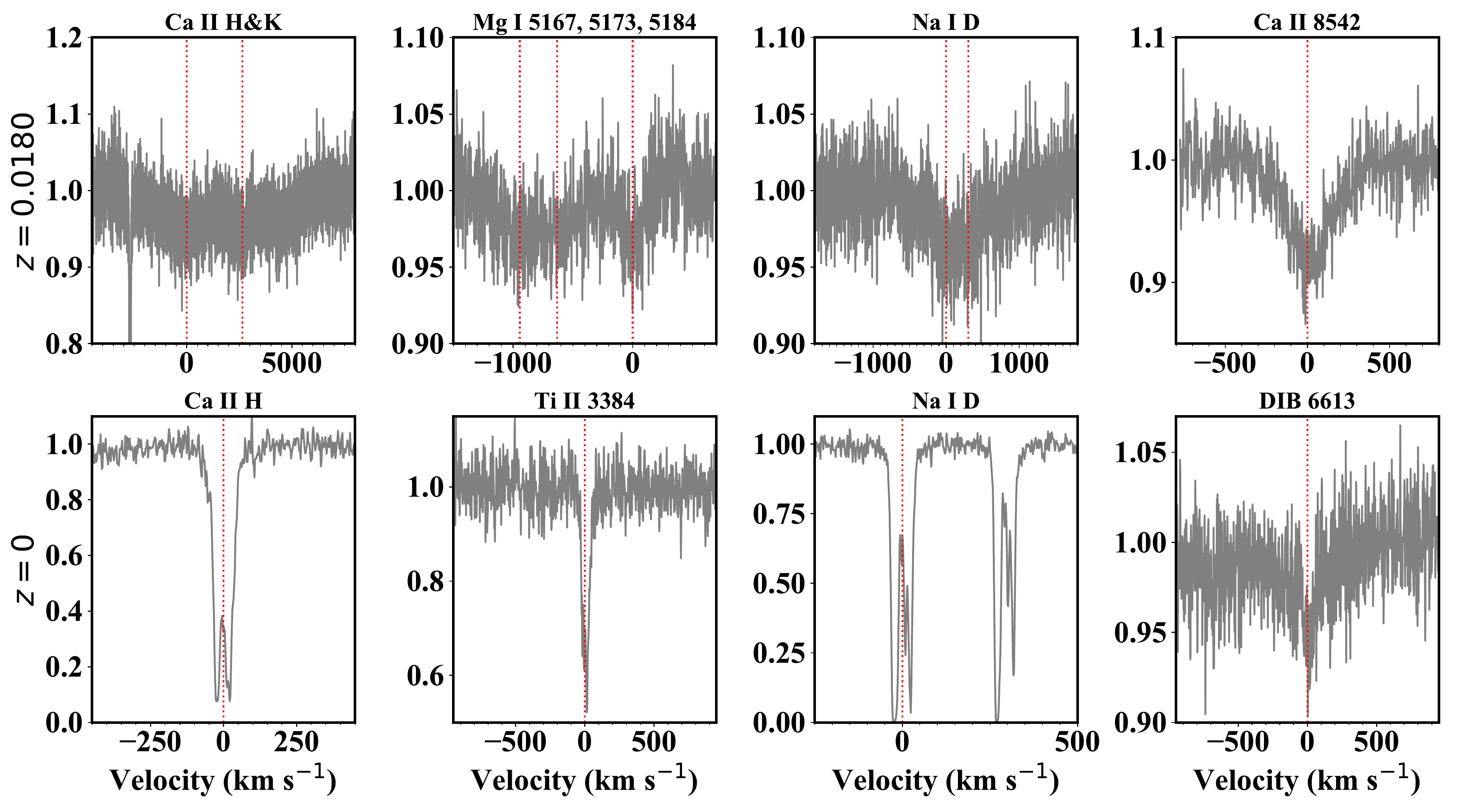}
\caption{A selection of narrow features from the UVES spectra, plotted in velocity space. The top panels show features at the host galaxy of AT~2018dyb, while the bottom panels show features at $z=0$. 
\label{fig:UVES}}
\end{figure*}

Especially interesting is the ratio of \ion{He}{2}/\ion{N}{3} $\lambda$4640. For both AT~2018dyb and ASASSN-14li,  \ion{N}{3} starts as the strongest line. However, this ratio evolves smoothly and \ion{He}{2} increases with time with respect to \ion{N}{3}. For ASASSN-14li the line ratio even reverses (as is also visible by eye). Interestingly, iPTF16fnl \citep{Onori16fnl} shows the opposite behaviour for this ratio, with \ion{He}{2}/\ion{N}{3} decreasing over time.

%%%%%%%%%%%%%%%%%%%%%%%%%%%%%%%%%%  RESULTS   Continued %%%%%%%%%%%%%%%%%%%%%%%%%%%%%%

\section{Early-time High-resolution Spectroscopy} 
\label{sec:highres}

We have used our UVES spectroscopy to search for narrow (of the order of 10--100 km~s$^{-1}$) absorption features in the spectra. We identify two kinds of lines: i) absorption and diffuse interstellar bands (DIBs) in the Milky Way at $z=0$ (the \ion{Na}{1} $\lambda\lambda$3302, 3303 UV doublet, \ion{Na}{1}~D, a couple of \ion{Ti}{2} transitions, \ion{Ca}{2} H and K, and several DIBs at 6196, 6204, 6284, and  6613 \AA); ii) lines originating at the host galaxy (\ion{Ca}{2} H and K, \ion{Mg}{1} $\lambda\lambda$5167, 5173, 5184, \ion{Na}{1}~D and two \ion{Ca}{2} lines in the NIR - the third falls in the UVES red CCD chip gap). Apart from these features, we do not detect any other line. In particular, we do not detect any line that we can attribute to the TDE itself, e.g. narrow blueshifted lines or P-Cygni absorption profiles that could be indicative of winds or outflowing material. 
A selection of narrow features from the UVES spectra are shown in Figure~\ref{fig:UVES}.

We have fit the NIR \ion{Ca}{2} lines to derive a redshift for the host of $z=0.0180$. 
These lines are not `narrow' but have width of a few hundred km~s$^{-1}$. In particular, we get an FWHM of $\sim290$ km~s$^{-1}$ for both Ca NIR lines (fit simultaneously) for both epochs. This indicates that they are not lines from the interstellar medium but stellar features at the host. 
\cite{Blago16fnl} used these lines to calculate the velocity dispersion at the host of iPTF16fnl and from there the mass of the supermassive black hole. 
However, \cite{Wevers2017} showed that an estimate based solely on these lines can give different results than a full template fitting for the determination of $\sigma$, probably because these lines can get collisionally broadened. 
The other host lines are quite shallow, but a forced fit on the \ion{Na}{1} D lines gives a similar width of $266 \pm 22$ km~s$^{-1}$. 
The host \ion{Ca}{2}~H and K lines are more difficult to study in the UVES spectrum because they are in a region with many strong TDE lines and the continuum determination is more ambiguous. 

The Galactic \ion{Na}{1}~D lines have a complex profile with four components and equivalent widths of 1.01~\AA\ for D1 and 0.78~\AA\ for D2. However, the lines are saturated. 
In addition, we measure an equivalent width of 0.076~\AA\ for the DIB at 6613 \AA, which is the strongest and `cleanest' among the DIBs detected. We note that the strength of these lines is perhaps larger than what is expected from the Galactic extinction in this direction. By using a relation provided by \cite{2011ApJ...727...33F}, the measured strength of the 6613 \AA\ DIB  would correspond to $E(B-V) \sim 0.37$~mag, which is almost double from the $E(B-V) \sim 0.2$~mag obtained by using the maps of \cite{2011ApJ...737..103S}. Nevertheless, these scaling relations are known to have significant scatter and we therefore  adopt the value of  \cite{2011ApJ...737..103S}.

%%%%%%%%%%%%%%%%%%%%%%%%%%%%%%%%%%  RESULTS   Continued %%%%%%%%%%%%%%%%%%%%%%%%%%%%%%

\section{Late-time Medium-resolution Spectroscopy and Black Hole Mass} 
\label{sec:medres}

\begin{figure*}
\includegraphics[width=\textwidth]{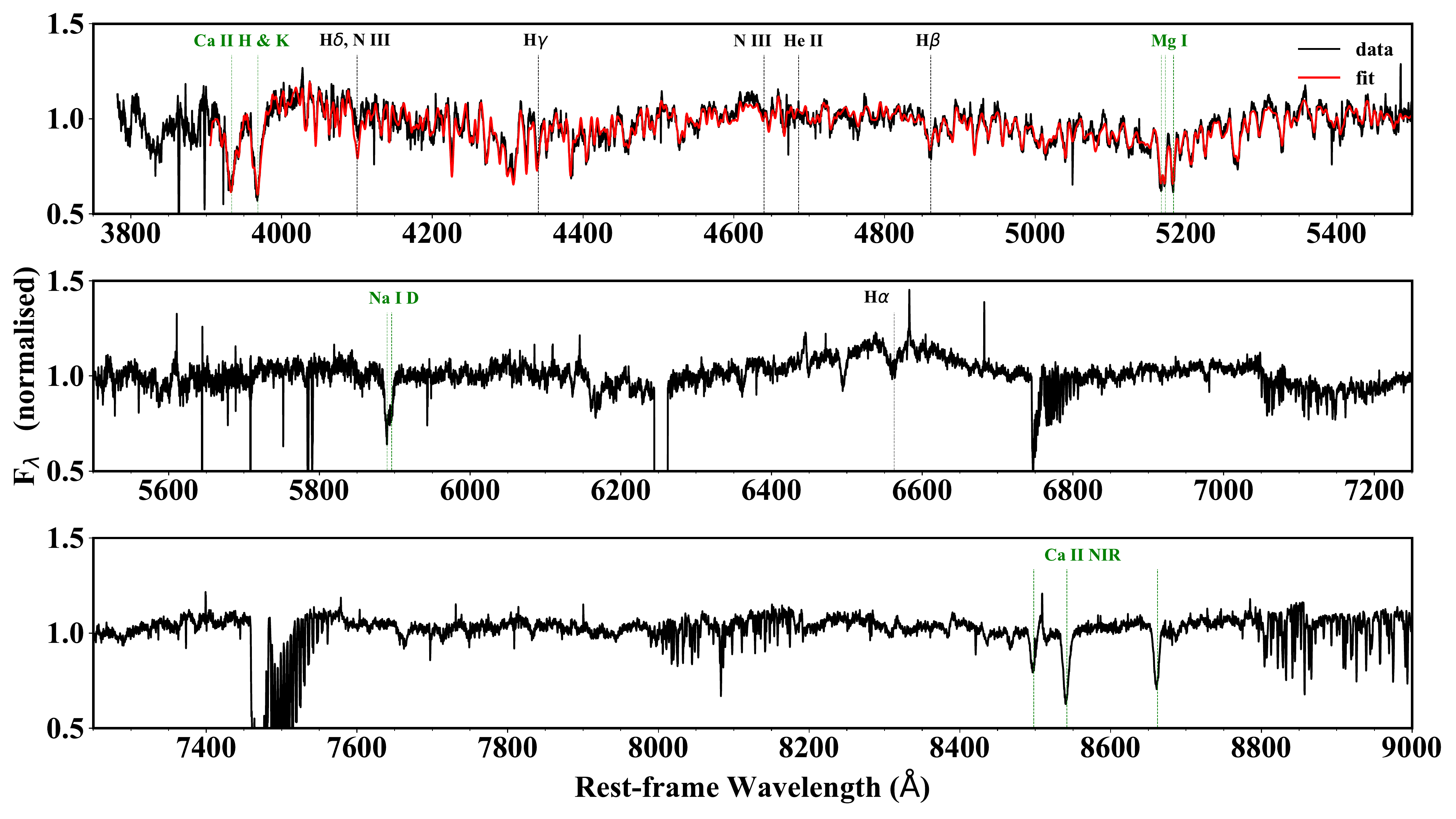}
\caption{The X-shooter spectrum of AT~2018dyb obtained 206 days after maximum light (normalised). The UV part has been fit by \textsc{ppxf} to determine the velocity dispersion at the host. The same lines have been marked as in previous figures for comparison: broad lines with black and narrow absorption lines with green dashed lines. With the exception of H$\alpha$, the broad lines that were prominent at early phases have disappeared.
\label{fig:XSHspectrum}}
\end{figure*}

The late-time spectra of AT~2018dyb are relatively free of TDE lines, except from H$\alpha$, 
and it is therefore easier to study the host galaxy and determine the black hole mass. 
Figure \ref{fig:XSHspectrum} shows the X-shooter spectrum, obtained 206 days after peak, normalised to the continuum. 
We have used \textsc{ppxf} \citep{Cappellari2017} in combination with the Elodie stellar template library \citep{Prugniel2001, Prugniel2007} to measure the velocity dispersion using the myriad of stellar absorption features present in the UVB arm of the spectrum (see \citealt{Wevers2017, Wevers2019} for more details). 
The best-fit template is shown in Figure \ref{fig:XSHspectrum}.
We thus measure a stellar velocity dispersion of 96 $\pm$ 1 km s$^{-1}$, corresponding to a black hole mass of $M_{\rm BH}$ = $3.3^{+5.0}_{-2.0}\times 10^6$ $M_{\odot}$ using the $M$--$\sigma$ relation of \citet{McConnellMa13}. 
The quoted uncertainty includes both the systematic and statistical uncertainties of the relation and measurement process, added linearly.

%%%%%%%%%%%%%%%%%%% RESULTS  Host galaxy and Light curve fits %%%%%%%%%%%%%%%%%%%%%%%%%%%%%%

\section{Host galaxy and host contamination}
\label{sec:host}

There are two main reasons to study the host galaxy of AT~2018dyb. The first is that
fundamental properties of the galaxy can be linked to the properties of the TDE progenitor system, i.e. the mass of the supermassive black hole \citep[e.g.][]{McConnellMa13} or the probable mass of the disrupted star \citep{Kochanek2016}. The second reason is that we are interested in placing constraints on the degree of host contamination of the light curves and spectra of AT~2018dyb.

To this end we modeled the host spectral energy distribution (SED) with the software package \texttt{LePhare}, version 2.2
\citep{Arnouts1999a,Ilbert2006a} \footnote{\href{http://www.cfht.hawaii.edu/~arnouts/LEPHARE/lephare.html}{http://www.cfht.hawaii.edu/\~{}arnouts/LEPHARE/lephare.html}}. 
For a detailed description of the SED modeling we refer to previous publications where a very similar procedure was used \citep{2011A&A...534A.108K,2016NatAs...1E...2L,2018MNRAS.473.1258S} and the references therein. As input, we used the archival $ugriz$ data from SkyMapper and the $JHK$ data from 2MASS. We performed photometry on images with equal seeing ($\sim$2.5$\arcsec$) with a small aperture (3$\arcsec$) in order to avoid the contamination by neighbouring objects. However, due to the large PSF and relatively large pixel scales, this may not have been entirely possible. For this reason we experimented with different combinations of filters and apertures and studied how this affected the SED modeling. We obtain a total stellar mass of $\log _{10}(M_{\star}/M_\odot)=10.08^{+0.25}_{-0.24}$ 
(where the central value is the median of the probability distribution and the error bars contain the $1\sigma$ probability interval). The stellar mass is relatively robust and does not seem to depend on the exact choice of input.

We observe that the light curves of AT~2018dyb, flattened after approximately 230 days past peak in all filters (Figure~\ref{fig:UVOT}).
The most natural explanation for this flattening is that the TDE has faded and that we are only measuring light from the host galaxy. This is not certain, however, and it should be verified with observations at even later phases, because \cite{2019ApJ...878...82V} have shown that TDEs often demonstrate late-time excess in the form of flattening. It is therefore possible that we have not yet reached the host level. Having this caveat in mind, it is possible to derive host magnitudes in all UVOT filters  by averaging the light curves after day $+230$. We derive the following magnitudes: $V$ = 16.63,  $B$ = 17.55, $U$ = 18.73, 
UVW1 = 19.42, UVM2 = 20.00 and UVW2 = 19.88~mag (all in the AB system). 
With these host magnitudes at hand, it is possible to estimate the different degrees of host contribution to the light curves at different times and at different wavelengths. We deduce that: 
i) the host contamination is negligible in the UV bands around maximum light; ii) the host contamination for UVW2, UVM2, and UVW1 stays always below 10\% up to day $+52$ and below 20\% up to day $+90$; iii) in the $U$ band the contamination increases to 30\% at day $+90$; iv) in the $B$ and $V$ bands the host contribution is very significant and it is never below 10\%; v) after day $+160$ (when AT~2018dyb reappeared behind the Sun), all light curves in Figure~\ref{fig:UVOT} were contaminated by host galaxy light by more than 50\%.

\section{Fits to the light curve}
\label{sec:mosfit}

\begin{figure}
\includegraphics[width=\columnwidth]{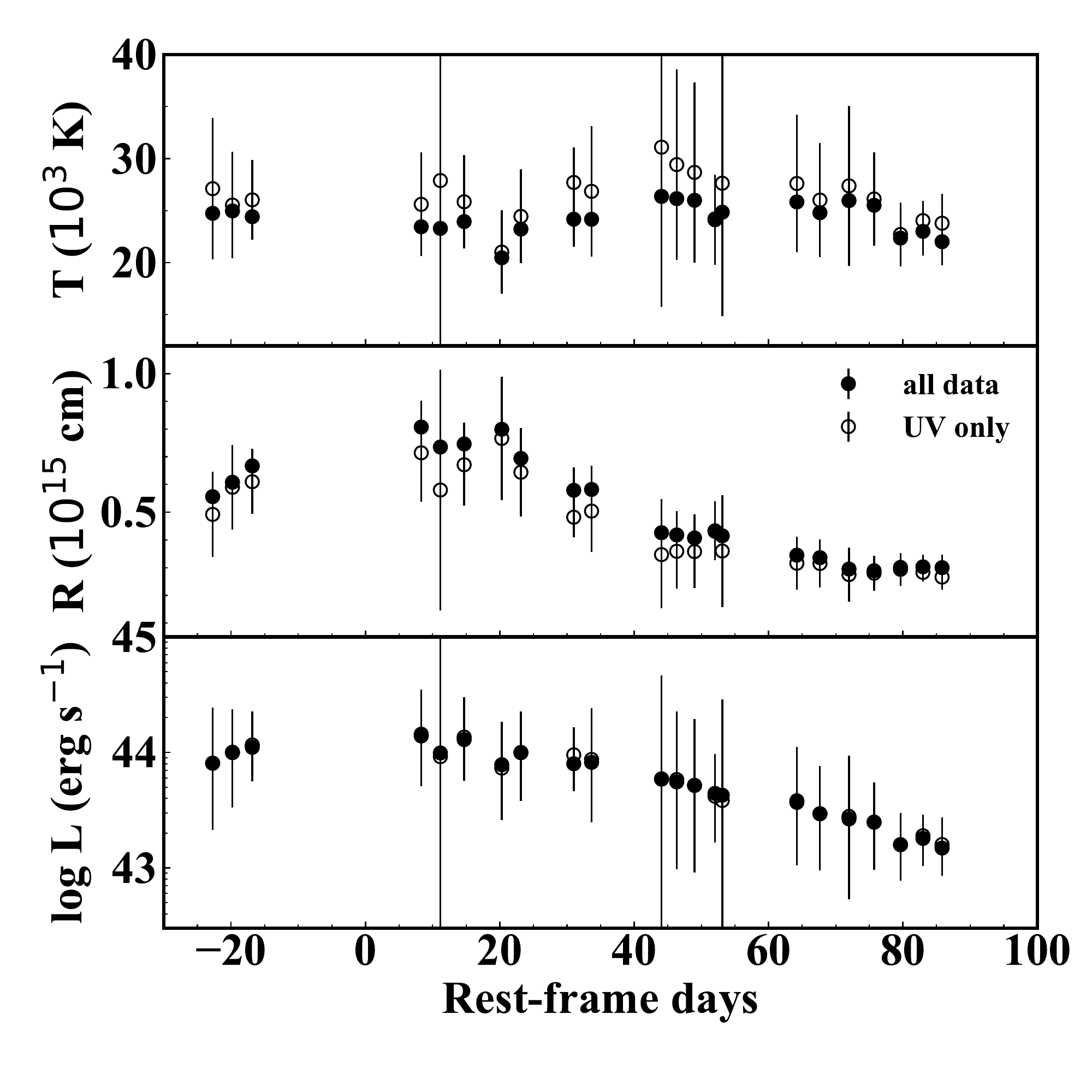}
\caption{Evolution of temperature, radius, and bolometric luminosity for AT~2018dyb, by fitting a blackbody to the UVOT photometry (after removal of the host contribution). We made two different fits, one including all UVOT data and one excluding the $B$ and $V$ bands. 
\label{fig:TRL}}
\end{figure}

By fitting a blackbody to the UVOT photometry of AT~2018dyb (after removal of our best estimate for host contamination; Section~\ref{sec:host}) we estimate the photospheric temperature,  blackbody radius and bolometric luminosity at different phases (Figure~\ref{fig:TRL}). We only fit data out to 100 days past maximum and we made two separate fits: one including all UVOT data and one where we excluded the $B$ and $V$ data.  
In both cases, we get consistent results, albeit with larger errors for the fit with UV-band data only. 
The temperature remains approximately constant around $\sim 25\,000$~K. In that sense, AT~2018dyb is similar to many other optical TDEs \citep[e.g.][]{Holoien14li,Hung16axa} although it is now documented that the temperature evolution of TDEs is quite diverse
\citep{Holoien15oi,Holoien18kh,2016NatAs...1E...2L,2019MNRAS.488.4816W}.

We used the transient fitting code {\tt{MOSFiT}}  \citep{Guillochon17} to fit the light curves of AT~2018dyb. {\tt{MOSFiT}} uses a library of  simulations of tidal disruptions from \cite{Guillochon2013} to calculate the mass fallback rate, and then scales the fallback rate using the properties of the black hole and disrupted star. It then converts the fallback rate into a bolometric luminosity curve and passes it through a viscous transform, approximating a viscous delay from an accretion disk or a diffusion delay through a dense photosphere. Finally, it uses a time-dependent reprocessing function to produce optical and UV light curves. The results of the code include best-fit parameter estimates for the masses of the black hole and the disrupted star. 
The {\tt{MOSFiT}}  TDE model remains agnostic as to whether the majority of the luminosity comes from accretion onto the black hole or from stream--stream collisions. The estimation of black hole mass relies largely on the relation between the peak timescale of TDEs and the mass of the disrupting black holes. In addition to the statistical errors from the model's Markov Chain Monte Carlo (MCMC) fit (e.g. the MCMC parameters account for potential changes to $t_{\rm peak}$ from slow circularization or inefficient accretion), the {\tt{MOSFiT}} model's error estimate accounts for potential systematic errors from uncertainty in the mass-radius relation of the disrupted star  \cite[see][for a more detailed analysis]{Mockler19}. This is likely to be the largest source of potential error in the peak timescale and therefore the estimated black hole mass that is not accounted for in the current {\tt{MOSFiT}} model parameters. Other factors that can affect the estimation of peak timescale and black hole mass include the spin of the star and the black hole, and very deep impact parameters. However, the magnitude of these effects will be less than the systematic uncertainty from the mass-radius relation unless the star is spinning near its break-up velocity \citep[$>0.2\times\Omega_{\rm breakup}$;][]{2019ApJ...872..163G}, 
or the impact parameter is very high \citep[$\beta > 6$;][]{2019MNRAS.487.4790G}. 
Highly spinning stars and very deep encounters are uncommon \citep{2016MNRAS.455..859S}, and {\tt{MOSFiT}} does not currently include their effects on the resultant light curves or the model error estimates.

\begin{figure}
\includegraphics[width=\columnwidth]{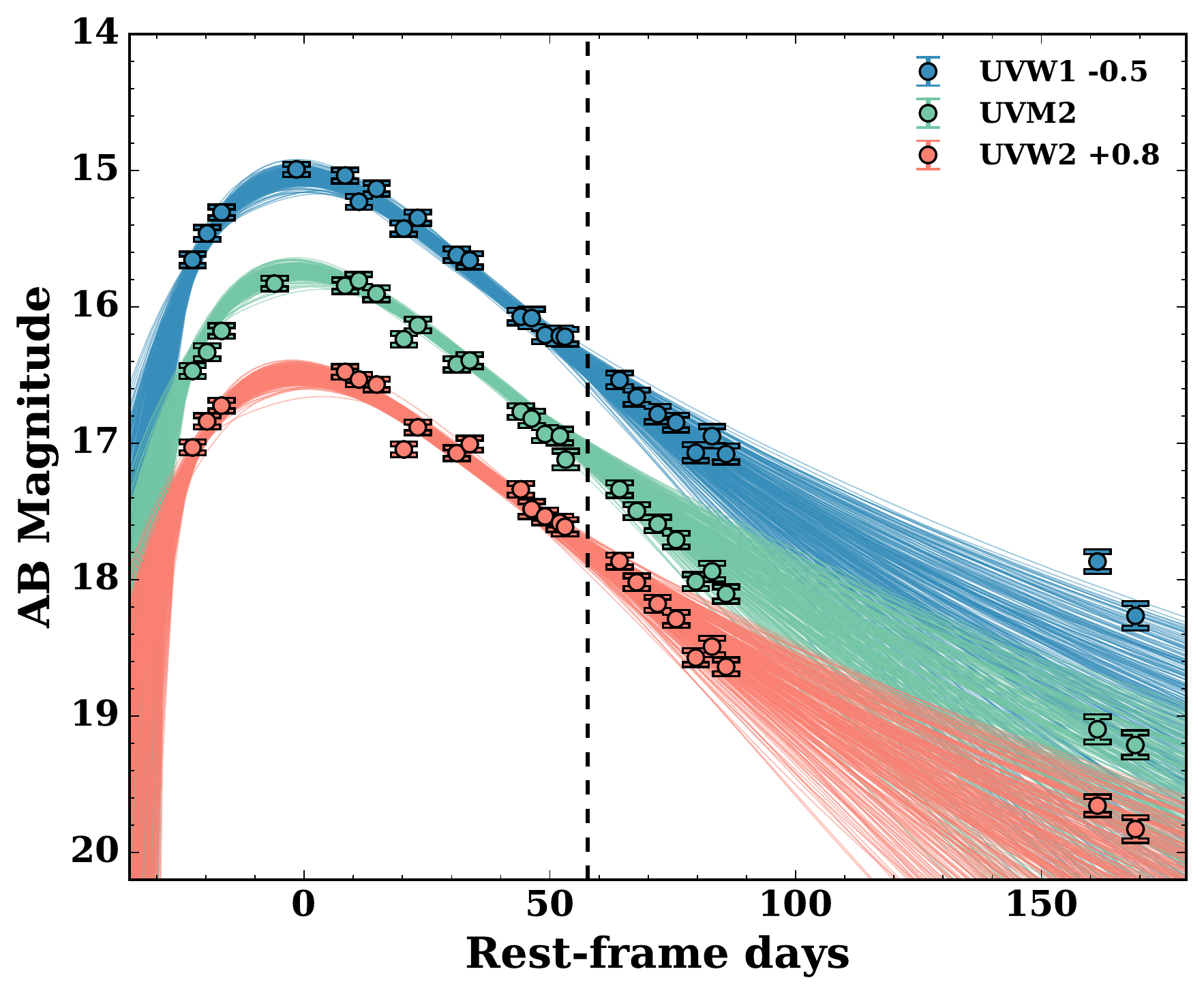}
\caption{Best-fit light curves from {\tt MOSFiT}. 
The fit only uses UV observations out to the dashed line at rest-frame day $+58.2$ ($\rm{MJD} = 58400$). 
As shown in Section~\ref{sec:host}, host contamination is negligible for these data and all the light comes from the TDE. 
The light curves have been extrapolated out to later times for the plot. 
\label{fig:MOSFiT}}
\end{figure}

The fits obtained for AT~2018dyb can be seen in Figure~\ref{fig:MOSFiT}. 
In this run we only fit data that are completely free of host contamination, i.e. only the bluest UV bands and only up to 55 days past peak (see Section~\ref{sec:host}).
The data have only been corrected  for Galactic extinction.  
We find that the best fit for the black hole mass is $4^{+5}_{-2}\times 10^6$  $M_{\odot}$, consistent with what we obtained by the $M$--$\sigma$ relation.
The best fit for the stellar mass is $0.7^{+4.0}_{-0.6}$ $M_{\odot}$. The fit preferred a full disruption of the star. The error bars include $\pm 0.2$ dex of systematic error in the black hole mass measurement and $\pm 0.66$ dex of systematic error in the stellar mass measurement.

%%%%%%%%%%%%%%%%%%%%%%%%%%%%%%%%%%  DISCUSSION   %%%%%%%%%%%%%%%%%%%%%%%%%%%%%%

\section{Discussion} 
\label{sec:disc}

The main observational result of this paper is the presence of strong N and O lines in the optical spectra of AT~2018dyb and the ascertainment that these lines are relatively common in optical TDEs.

TDEs can have N features due to its enhancement in the debris falling back at some phase after peak \citep{kochanek16}. This is especially the case for the disruption of more massive stars since their N abundance increases more dramatically over the main-sequence evolution \citep{Gallegos-Garcia18}. However, the enhancement in the TDE debris fallback itself is not sufficient to explain the strong N line observed here with flux comparable to H$\alpha$ and \ion{He}{2}. For example, even for a massive star of mass $\sim 3M_\odot$ near the end of its main-sequence evolution, the fallback abundance of N can only be enhanced to eight times of its solar abundance, still putting it at a ratio of $\sim1:100$ compared with H. 
Furthermore, from the  {\tt{MOSFiT}} fit the disrupted star in AT 2018dyb is more likely to have a low mass, so the N enhancement over its lifetime should not be very significant. Therefore, we conclude that Bowen fluorescence is the most likely mechanism for producing the observed strong \ion{N}{3} features by exciting N with energy unavailable to H and He.

It is reassuring that so far all TDEs with prominent Bowen features (AT 2018dyb, ASASSN14-li, iPTF15af, iPTF16axa, iPTF16fnl) are all accompanied by strong \ion{He}{2} optical lines, since Bowen fluorescence is primarily triggered by the ionization and recombination of \ion{He}{2}. The detection of \ion{O}{3} lines further lends support to this mechanism. 
However, not all TDEs with \ion{He}{2} lines have strong Bowen features, perhaps because they do not produce the optimal physical conditions (such as temperature, optical depth, and velocity gradient) needed for strong resonance to happen \citep{Weymann69, Kallman80, Selvelli07}. 
It is also possible that the nitrogen fallback time can affect the phases where Bowen features can be observed.
Interestingly, all of the hitherto identified N-rich TDEs also show Balmer emission besides \ion{He}{2}. In addition, for both AT~2018dyb and ASASSN-14li, the ratio of \ion{N}{3} to \ion{He}{2} is observed to decrease with time. The exact ratio of the different emission lines and their evolution (Figure~\ref{fig:lineevol2018dyb}) can be used to shed light on the physical conditions in a TDE. For this, however, detailed modeling of the radiative transfer physics \cite[e.g.][]{1985ApJ...299..752N} is required, which is beyond the scope of this paper.

In any case, the detection of prominent Bowen fluorescent lines in TDEs has important implications for the mechanism of emission and the geometry of the emitting gas.

TDEs with Bowen lines should also produce strong EUV or X-ray emission, since a large flux of photons with wavelength shorter than 228 {\AA} is needed to ionize \ion{He}{2} and trigger the Bowen mechanism. 
This flux cannot be provided by the blue tail of the UV/optical blackbody continuum that we observe. By integrating a blackbody of $25\,000$~K, we find that only $\sim 10^{-8}$ of the total luminosity is emitted below 228~\AA, corresponding to a luminosity of the order of $10^{36}$~erg~s$^{-1}$ (Figure~\ref{fig:TRL}). 
In comparison, the observed Bowen lines have line luminosities of the order of $10^{41}$~erg~s$^{-1}$ (Table~\ref{tab:linelum}), so another ionising source is required.

X-ray/EUV emission in TDEs is usually associated with the accretion process of the transient debris disk \citep{Cannizzo90, Ulmer99, Lodato11}, since the collisions of debris streams near the apocenter of their orbits will mostly produce optical photons \citep{Piran15} unless the star has penetrated very deeply into the tidal disruption radius. In the case of a deep plunge, however, a prompt formation of an accretion disk is also expected due to efficient removal oforbital energy  in debris stream collisions \citep{Dai15, Guillochon15}. Therefore, the indirect uncovering of EUV/X-ray photons through the detection of Bowen lines in the optical spectrum favors the accretion paradigm for optical TDEs.

This unobserved EUV radiation energy component may also (partially) solve the ``TDE missing energy problem'' \citep{Lu18}, which states that the observed radiation energy in the optical/UV bands for optical TDEs is less than $\sim10\%$ of the rest-mass energy of a star. 

EUV radiation is hard to detect due to Galactic extinction. Strong X-rays, however, are also lacking in most optical TDEs with Bowen features except in ASASSN-14li \citep{Miller15, Holoien14li}. This, however, can be a geometric effect as X-ray photons produced by the inner disk are obscured by some gaseous medium such as the unbound debris stream or by winds \citep{Loeb97, Strubbe09, Coughlin14, Guillochon14}. Since the debris stream is confined by gravity and therefore has a vertically thin structure \citep{Kochanek94,Guillochon14}, it is unlikely that it will cover a large solid angle. This leaves obscuration by wind as the most likely explanation. 
It has been proposed that optical TDEs can be produced by the reprocessing of X-ray photons in optically thick winds formed in TDEs \citep{Metzger16, Roth16}. Furthermore, \citet{Dai18} employed a super-Eddington TDE disk simulation to show that optically thick winds are indeed produced from such disks. Based on the anisotropic wind profile, a unified picture of X-ray to optical TDEs can be provided by viewing the disk from different inclinations. 
The optical depth of the super-Eddington disk and wind is large, allowing for multiple scatterings to happen before photons leak out of the system. This can greatly boost the resonance of \ion{O}{3} and \ion{N}{3} at 304 {\AA} and increase the efficiency of the Bowen mechanism. Most of the radiation energy, indeed, is emitted in the EUV range in this model. 

In the electron scattering-dominated regime, the emission lines should be broadened by scattering and therefore the line width is primarily set by the number of scatterings of photons (or the optical depth) instead of gas kinematics \citep{Roth18}. The narrowing of the lines observed in AT~2018dyb (and other TDEs) after peak is consistent with the decrease in optical depth of the system as the debris fallback rate and accretion rate drop with time.
In the unified TDE model of \cite{Dai18}, the electron scattering photosphere is larger in the disk direction and tucks in near the pole \citep[see also][]{2019MNRAS.488.1878N}. 
This can explain why, between the two TDEs with the strongest Bowen features, ASASSN-14li has narrower lines than AT~2018dyb: ASASSN-14li is viewed at a smaller inclination angle to the pole (consistent with the detection of X-rays), while AT~2018dyb is viewed closer to the disk (larger inclination angle). This makes AT~2018dyb opaque to X-rays, while at the same time the photons need to go through more scattering to leak out of the photosphere, resulting in broader line profiles.

%%%%%%%%%%%%%%%%%%%%%%%%%%%%%%%%%%  CONCLUSION   %%%%%%%%%%%%%%%%%%%%%%%%%%%%%%

\section{Conclusions} \label{sec:conc}

The main points of this study can be summarized as follows.
\begin{itemize}
\item We have unambiguously detected strong lines of N and O in the optical spectra of AT~2018dyb (ASASSN-18pg). 
\item We have shown that these lines are quite common in the  spectra of TDEs and that there exist optical TDEs that are `N-rich'. 
\item The \ion{N}{3} and \ion{O}{3} lines most likely originate from the Bowen fluorescence mechanism.
\item The detection of the Bowen lines requires the existence of EUV/X-ray photons, and this argues for an accretion origin for optical TDEs.
\item The strongest emission lines appear slightly blueshifted at early times but progressively move to the red. 
\item The FWHM of the emission lines decreases with time. This is expected if their width is primarily set by electron scattering \citep{Roth18} and if the optical depth decreases with time. 
\item \ion{N}{3} starts stronger than \ion{He}{2} but the ratio \ion{He}{2} / \ion{N}{3} increases with time. This is possibly associated with the change in the ionization level. 
\item These last observations are also valid for ASASSN-14li, which has a spectrum very similar to AT~2018dyb. Two important differences are that the lines of ASASSN-14li are narrower by a factor of $\sim$2 and that it is detected in X-rays. 
\item These differences can be explained within the unification model of \cite{Dai18} where TDEs are observed through a super-Eddington disk with an  anisotropic optically thick wind. Within this scenario, AT~2018dyb is viewed at a larger polar inclination (closer to the disk mid-plane) than ASASSN-14li and this is why it is opaque to X-rays and why its lines become broader, because they go through more scattering on the way out of the photosphere.
\item High-resolution spectroscopy of AT~2018dyb obtained close to peak does not reveal any narrow features that can be safely attributed to the TDE (e.g. due to outflows).
\item The host galaxy mass is found to be $\log _{10}(M_{\star}/M_\odot)=10.08^{+0.25}_{-0.24}$ by SED fitting. 
\item The black hole mass is  estimated by the $M$--$\sigma$ relation to be $M_{\rm BH}$ = $3.3^{+5.0}_{-2.0}\times 10^6$ $M_{\odot}$.
\item A consistent estimate is obtained by using \texttt{MOSFiT} that yields $4^{+5}_{-2}\times 10^6$  $M_{\odot}$. In addition, this fit  predicts that AT~2018dyb was produced by the disruption of a $0.7^{+4.0}_{-0.6}$ $M_{\odot}$ star. 

\end{itemize}

The observations of AT~2018dyb constitute an excellent dataset that can be used to understand the physical conditions in TDEs through detailed modeling.

%%%%%%%%%%%%%%%%%%%%%%%%%%%%%%%%%%  THE REST   %%%%%%%%%%%%%%%%%%%%%%%%%%%%%%

\begin{acknowledgements}
 
GL and DBM are supported by a research grant (19054) from VILLUM FONDEN. 
The Cosmic Dawn Centre is supported by the DNRF. 
LD and ER acknowledge the support from DNRF.
IA is a CIFAR Azrieli Global Scholar in the Gravity and the Extreme Universe Program and acknowledges support from that program, from the Israel Science Foundation (grant numbers 2108/18 and 2752/19), from the United States - Israel Binational Science Foundation (BSF), and from the Israeli Council for Higher Education Alon Fellowship.
PGJ, GC and ZKR acknowledge support from European Research Council Consolidator Grant 647208.
KM acknowledges support from STFC (ST/M005348/1) and from H2020 through an ERC Starting Grant (758638).
MG was supported by the Polish National Science Centre grant OPUS 2015/17/B/ST9/03167.
T.-W.C. acknowledges funding from the Alexander von Humboldt Foundation.
SJS acknowledges funding from STFC Grants  ST/P000312/1 and ST/N002520/1. 
This work is based on observations collected at the European Organisation for Astronomical Research in the Southern Hemisphere, Chile as part of ePESSTO, (the Public ESO Spectroscopic Survey for Transient Objects Survey) ESO program 199.D-0143.
MG is supported by the Polish NCN MAESTRO grant 2014/14/A/ST9/00121.

\end{acknowledgements}

%%%%%%%%%%%%%%%%%%%%%%%%%%%%%%%%%%%%%%%%%%%%%%%%%%%%%%%%%%%%%%%%%%%%%%

%\bibliographystyle{apj} 
%\bibliography{AT2018dyb_specevol.bib}

\bibliographystyle{apj}

%%%%%%%%%%%%%%%%%%%%%%%%%%%%%%%%%%%%%%%%%%%%%%%%%%%%%%%%%%%%%%%%%%%%%%

\clearpage
\begin{deluxetable}{ccrcccccc}
  %\tabletypesize{\scriptsize}
  %\rotate
  \tablecaption{UVOT photometry of AT~2018dyb \tablenotemark{a} \label{tab:UVOTphot}}
  \tablehead{
      \colhead{UT date} & 
    \colhead{MJD} & 
    \colhead{Phase \tablenotemark{b}} & 
    \colhead{UVW2} &
    \colhead{UVM2} &
    \colhead{UVW1} & 
    \colhead{$U$}  &
    \colhead{$B$} & 
    \colhead{$V$} \\
    (yy-mm-dd)     &    (days)      &     (days)      &     (mag)     &       (mag)     &      (mag)      &       (mag)     &       (mag)     &    (mag)            }
  \startdata
2018-07-18 &   58317.63 & $-$22.7 &  16.28 (0.07) &  16.55 (0.06) &  16.25 (0.06) &  15.92 (0.05) &  15.81 (0.05) &  15.59 (0.05) \\ 
2018-07-21 &   58320.65 & $-$19.7 &  16.08 (0.07) &  16.38 (0.07) &  16.00 (0.06) &  15.69 (0.05) &  15.65 (0.05) &  15.57 (0.06) \\ 
2018-07-24 &   58323.64 & $-$16.8 &  15.97 (0.07) &  16.25 (0.06) &  15.86 (0.06) &  15.61 (0.05) &  15.50 (0.05) &  15.45 (0.05) \\ 
2018-08-04 &   58334.61 &  $-$6.0 &        \nodata   &  15.91 (0.06) &        \nodata   &        \nodata   &        \nodata   &        \nodata   \\ 
2018-08-06 &   58336.67 & $-$4.0 &        \nodata   &        \nodata   &        \nodata   &  15.34 (0.05) &        \nodata   &        \nodata   \\ 
2018-08-09 &   58339.15 &  $-$1.6 &        \nodata   &        \nodata   &  15.56 (0.06) &        \nodata   &        \nodata   &        \nodata   \\ 
2018-08-19 &   58349.24 &     8.4 &  15.71 (0.07) &  15.90 (0.06) &  15.62 (0.06) &  15.32 (0.05) &  15.21 (0.05) &  15.17 (0.05) \\ 
2018-08-22 &   58352.09 &    11.2 &  15.82 (0.07) &  15.91 (0.06) &  15.99 (0.06) &  15.45 (0.05) &  15.37 (0.05) &  15.29 (0.06) \\ 
2018-08-25 &   58355.74 &    14.7 &  15.83 (0.07) &  15.99 (0.06) &  15.70 (0.06) &  15.45 (0.05) &  15.35 (0.05) &  15.24 (0.05) \\ 
2018-08-31 &   58361.39 &    20.3 &  16.32 (0.07) &  16.32 (0.06) &  16.03 (0.06) &  15.61 (0.05) &  15.53 (0.05) &  15.37 (0.05) \\ 
2018-09-03 &   58364.28 &    23.1 &  16.15 (0.07) &  16.19 (0.06) &  15.91 (0.06) &  15.67 (0.05) &  15.52 (0.05) &  15.44 (0.05) \\ 
2018-09-11 &   58372.35 &    31.0 &  16.32 (0.07) &  16.52 (0.06) &  16.17 (0.06) &  15.98 (0.06) &  15.72 (0.05) &  15.46 (0.06) \\ 
2018-09-14 &   58375.03 &    33.7 &  16.28 (0.07) &  16.51 (0.06) &  16.22 (0.07) &  15.91 (0.06) &  15.75 (0.07) &  15.55 (0.08) \\ 
2018-09-24 &   58385.61 &    44.1 &  16.60 (0.07) &  16.84 (0.06) &  16.71 (0.06) &  16.30 (0.06) &  16.04 (0.06) &  15.89 (0.07) \\ 
2018-09-26 &   58387.86 &    46.3 &  16.76 (0.08) &  16.91 (0.07) &  16.71 (0.08) &  16.41 (0.08) &  16.10 (0.09) &  15.87 (0.12) \\ 
2018-09-29 &   58390.63 &    49.0 &  16.81 (0.08) &  17.00 (0.07) &  16.76 (0.07) &  16.43 (0.07) &  16.13 (0.07) &  16.00 (0.09) \\ 
2018-10-02 &   58393.71 &    52.0 &  16.99 (0.08) &  17.09 (0.07) &  16.77 (0.07) &  16.45 (0.07) &  16.26 (0.08) &  16.13 (0.11) \\ 
2018-10-03 &   58394.77 &    53.1 &  16.86 (0.08) &        \nodata   &  16.81 (0.07) &  16.49 (0.07) &  16.20 (0.07) &        \nodata   \\ 
2018-10-03 &   58394.90 &    53.2  &        \nodata   &  17.18 (0.08) &        \nodata   &        \nodata   &        \nodata   &    15.89 (0.13) \\ 
2018-10-15 &   58406.06 &    64.2 &  17.15 (0.07) &  17.42 (0.06) &  17.10 (0.07) &  16.74 (0.06) &  16.48 (0.06) &  15.95 (0.07) \\ 
2018-10-18 &   58409.63 &    67.7 &  17.31 (0.08) &  17.61 (0.07) &  17.21 (0.07) &  16.85 (0.07) &  16.54 (0.07) &  16.14 (0.08) \\ 
2018-10-22 &   58413.96 &    71.9 &  17.44 (0.08) &  17.70 (0.07) &  17.38 (0.07) &  16.97 (0.07) &  16.67 (0.07) &  16.09 (0.08) \\ 
2018-10-26 &   58417.77 &    75.7 &  17.56 (0.08) &  17.80 (0.07) &  17.42 (0.07) &  17.06 (0.07) &  16.66 (0.07) &  16.41 (0.09) \\ 
2018-10-30 &   58421.85 &    79.7 &  17.88 (0.08) &  18.10 (0.07) &  17.62 (0.07) &  17.22 (0.07) &  16.83 (0.08) &  16.25 (0.09) \\ 
2018-11-03 &   58425.21 &    83.0 &  17.78 (0.09) &  18.05 (0.08) &  17.49 (0.08) &  17.23 (0.09) &  16.66 (0.09) &  16.44 (0.13) \\ 
2018-11-06 &   58428.13 &    85.8 &  17.91 (0.08) &  18.20 (0.07) &  17.66 (0.07) &  17.32 (0.08) &  16.74 (0.08) &  16.17 (0.09) \\ 
2019-10-22 &   58505.04 &   161.4 &  18.96 (0.09) &  19.18 (0.10) &  18.58 (0.09) &  18.08 (0.08) &  17.27 (0.07) &  16.49 (0.07) \\ 
2019-01-29 &   58512.90 &   169.1 &  19.10 (0.10) &  19.27 (0.10) &  18.83 (0.10) &  18.10 (0.09) &  17.27 (0.08) &  16.62 (0.11) \\ 
2019-02-02 &   58516.69 &   172.8 &  19.10 (0.10) &  19.25 (0.09) &  18.77 (0.10) &  18.33 (0.10) &  17.20 (0.08) &  16.55 (0.09) \\ 
2019-02-11 &   58525.20 &   181.2 &  19.36 (0.11) &  19.46 (0.11) &  18.87 (0.10) &  18.32 (0.10) &  17.48 (0.09) &  16.66 (0.10) \\ 
2019-02-14 &   58528.38 &   184.3 &  19.43 (0.10) &  19.49 (0.09) &  19.02 (0.10) &  18.28 (0.09) &  17.34 (0.08) &  16.51 (0.08) \\ 
2019-02-18 &   58532.49 &   188.4 &  19.37 (0.10) &  19.39 (0.12) &  18.85 (0.09) &  18.33 (0.09) &  17.34 (0.08) &  16.58 (0.08) \\ 
2019-02-26 &   58540.54 &   196.3 &  19.57 (0.12) &  19.47 (0.10) &  19.15 (0.12) &  18.41 (0.11) &  17.43 (0.09) &  16.60 (0.10) \\ 
2019-03-02 &   58544.23 &   199.9 &  19.66 (0.12) &  19.60 (0.10) &  19.17 (0.13) &  18.49 (0.13) &  17.56 (0.11) &  16.62 (0.11) \\ 
2019-03-06 &   58548.34 &   203.9 &  19.74 (0.13) &  19.73 (0.12) &  19.48 (0.15) &  18.36 (0.11) &  17.41 (0.10) &  16.71 (0.11) \\ 
2019-03-10 &   58552.63 &   208.1 &  19.62 (0.11) &  19.84 (0.11) &  19.28 (0.12) &  18.62 (0.12) &  17.52 (0.09) &  16.49 (0.09) \\ 
2019-03-19 &   58561.82 &   217.2 &  19.69 (0.11) &  20.14 (0.19) &  19.25 (0.14) &  18.63 (0.11) &        \nodata   &        \nodata   \\ 
2019-03-23 &   58565.43 &   220.7 &  19.93 (0.13) &  19.91 (0.16) &  19.24 (0.13) &  18.51 (0.10) &        \nodata   &        \nodata   \\ 
2019-03-31 &   58573.48 &   228.6 &  19.98 (0.12) &  20.09 (0.15) &  19.34 (0.12) &  18.66 (0.09) &        \nodata   &        \nodata   \\ 
2019-04-08 &   58581.27 &   236.3 &  19.86 (0.12) &  19.97 (0.15) &  19.49 (0.13) &  18.68 (0.10) &  17.47 (0.08) &  16.73 (0.08) \\ 
2019-04-11 &   58584.39 &   239.3 &  19.79 (0.13) &  20.11 (0.17) &  19.38 (0.14) &  18.82 (0.12) &  17.52 (0.09) &  16.66 (0.09) \\ 
2019-04-17 &   58590.91 &   245.7 &  19.73 (0.11) &  19.98 (0.15) &  19.48 (0.13) &  18.70 (0.10) &  17.54 (0.08) &  16.71 (0.08) \\ 
2019-04-20 &   58593.47 &   248.3 &  19.85 (0.11) &  19.95 (0.14) &  19.27 (0.11) &  18.62 (0.09) &        \nodata   &        \nodata   \\ 
2019-04-20 &   58593.96 &   248.7 &  19.73 (0.12) &  20.24 (0.17) &  19.48 (0.14) &  18.77 (0.11) &  17.54 (0.08) &  16.58 (0.08) \\ 
2019-04-23 &   58596.42 &   251.2 &  19.97 (0.12) &  19.91 (0.14) &  19.23 (0.12) &  18.71 (0.10) &  17.52 (0.08) &  16.56 (0.07) \\ 
2019-05-02 &   58605.67 &   260.2 &  19.82 (0.11) &  19.97 (0.14) &  19.35 (0.12) &  18.57 (0.09) &  17.53 (0.07) &  16.60 (0.07) \\ 
2019-05-05 &   58608.64 &   263.2 &  19.94 (0.12) &  20.14 (0.16) &  19.54 (0.13) &  18.70 (0.10) &  17.59 (0.08) &  16.64 (0.08) \\ 
2019-05-11 &   58614.74 &   269.2 &  19.78 (0.16) &        \nodata   &  19.32 (0.12) &  18.80 (0.10) &  17.54 (0.08) &        \nodata   \\ 
2019-05-14 &   58617.53 &   271.9 &  19.89 (0.12) &  19.97 (0.14) &  19.39 (0.13) &  18.80 (0.11) &  17.58 (0.08) &  16.67 (0.08) \\ 
2019-05-14 &   58617.69 &   272.1 &  19.98 (0.13) &  20.06 (0.16) &  19.18 (0.12) &  18.51 (0.10) &        \nodata   &        \nodata   \\ 
2019-05-17 &   58620.58 &   274.9 &  19.78 (0.12) &  19.91 (0.15) &  19.46 (0.13) &  18.67 (0.10) &  17.58 (0.08) &  16.53 (0.08) \\ 
2019-05-19 &   58622.71 &   277.0 &  19.79 (0.10) &  19.83 (0.12) &  19.38 (0.11) &  18.73 (0.09) &        \nodata   &        \nodata   \\ 
2019-05-20 &   58623.50 &   277.8 &  19.76 (0.12) &  19.69 (0.13) &  19.21 (0.12) &  18.85 (0.11) &  17.60 (0.08) &  16.61 (0.08) \\
2019-05-24 &   58627.60 &   281.8 &  19.81 (0.10) &  20.01 (0.13) &  19.59 (0.11) &  18.70 (0.08) &     \nodata     &        \nodata     \\ 
2019-05-29 &   58632.50 &   286.6 &  19.93 (0.10) &  19.98 (0.12) &  19.49 (0.11) &  18.78 (0.09) &     \nodata     &        \nodata     \\ 
2019-06-08 &   58642.54 &   296.5 &  20.12 (0.14) &  20.33 (0.20) &  19.53 (0.15) &  18.96 (0.13) &     \nodata     &        \nodata     \\ 
2019-06-14 &   58648.85 &   302.7 &  19.98 (0.12) &  19.93 (0.13) &  19.44 (0.12) &  18.70 (0.09) &     \nodata     &        \nodata     \\ 
2019-06-19 &   58653.52 &   307.3 &  19.95 (0.11) &  19.78 (0.12) &  19.44 (0.11) &  18.62 (0.08) &     \nodata     &        \nodata     \\ 
2019-06-23 &   58657.72 &   311.4 &  20.03 (0.12) &  20.22 (0.15) &  19.57 (0.12) &  18.85 (0.10) &     \nodata     &        \nodata     \\ 
2019-07-03 &   58667.66 &   321.1 &  19.96 (0.10) &  20.07 (0.13) &  19.51 (0.11) &  18.77 (0.09) &     \nodata     &        \nodata    
\enddata
\tablenotetext{a}{The magnitudes are given in the AB magnitude system and they are not corrected for Galactic extinction or host galaxy contamination. 
}
\tablenotetext{b}{With respect to the date of $U$-band maximum ($\mathrm{MJD} = 58340.74$) and given in the rest frame of AT~2018dyb ($z=0.0180$).}
\end{deluxetable}

\begin{deluxetable}{ccrcccc}
  \tablecaption{Log of spectroscopic observations \label{tab:speclog}}
 \tablehead{
      \colhead{UT date} & 
    \colhead{MJD} & 
    \colhead{Phase \tablenotemark{a}} & 
    \colhead{Telescope+Instrument} &
    \colhead{Grism/Grating} &
        \colhead{Slit Width} &
    \colhead{Exposure Wime} \\
    (yy-mm-dd)     &    (days)      &     (days)      &       &          &   (arcsec)       &  (s)   }
\startdata
2018-07-19 	&   58318.42 & $-$21.9 & LCO+FLOYDS 	& red/blu    			&  2	&   2700     \\
2018-07-22 	&   58321.22 & $-$19.2 & VLT+UVES 		& 346+580   			&  1	&  1800 \\
2018-07-23 	&   58322.21 & $-$18.2 & VLT+UVES 		& 346+580, 437+860   	&  1	&  1800, 1800 \\
2018-08-03 	&   58333.11 & $-$7.5 & NTT+EFOSC2 	& GR\#11, GR\#16   	  	&  1	&  1500, 1500 \\
2018-08-07 	&   58337.45 & $-$3.2 & LCO+FLOYDS 	& red/blu    			&  2	&   2700   \\
2018-08-13 	&   58343.13 &     2.3 & NTT+EFOSC2 	& GR\#11, GR\#16   		&  1	&  1500, 1500   \\
2018-08-13 	&   58343.47 &     2.7  & LCO+FLOYDS 	& red/blu    			&  2	&   2700   \\
2018-08-16 	&   58346.01 &     5.2  & VLT+UVES 		& 346+580, 437+860   	&  1	&  1800, 1800  \\
2018-08-18 	&   58348.14 &     7.3  & NTT+EFOSC2 	& GR\#11, GR\#16   		&  1	&  1200, 1200   \\
2018-08-28 	&   58358.41 &    17.4 & LCO+FLOYDS 	& red/blu    			&  2	&   2700   \\
2018-09-01 	&   58362.98 &    21.8 & NTT+EFOSC2 	& GR\#11, GR\#16   		&  1.5, 1	&  1800, 1800   \\
2018-09-15 	&   58376.03 &    34.7 & NTT+EFOSC2 	& GR\#11, GR\#16   		&  1	&  1800, 1800  \\
2018-09-16 	&   58377.46 &    36.1 & LCO+FLOYDS 	& red/blu    			&  2	&   2700   \\
2018-10-02 	&   58393.99 &    52.3 & NTT+EFOSC2 	& GR\#11, GR\#16   		&  1	&  1800, 1800   \\
2018-10-18 	&   58409.00 &    67.1 & NTT+EFOSC2 	& GR\#13   			&  1	&  1800   \\
2019-01-25 	&   58508.31 &   164.6 & NTT+EFOSC2 	& GR\#13   			&  1	&   2$\times$1800\\
2019-02-10 	&   58524.32 &   180.3 & NTT+EFOSC2 	& GR\#11   			&  1	&   2700 \\
2019-02-25 	&   58539.30 &   195.1 & NTT+EFOSC2 	& GR\#11, GR\#16      	&  1	&  2700, 2700      \\
2019-03-08 	&   58550.24 &  205.8  & VLT+X-shooter	 	& UVB, VIS, NIR		& 1, 1, 0.9 &  3680, 3680, 3840  \\
2019-03-17 	&   58559.34 &  214.7  & NTT+EFOSC2 	& GR\#11   			&  1	&   2700 \\
2019-05-01 	&   58604.29 &  258.9  & NTT+EFOSC2 	& GR\#11   			&  1	&   2700 \\
2019-07-30 	&   58694.13 &  347.1  & NTT+EFOSC2 	& GR\#11   			&  1	&   2700 
\enddata
\tablenotetext{a}{With respect to the date of $U$-band maximum ($\mathrm{MJD} = 58340.74$) and given in the rest frame of AT~2018dyb ($z=0.0180$).}
\end{deluxetable}

\begin{deluxetable}{rccccccc}
  \tablecaption{Emission line luminosities \tablenotemark{a} \label{tab:linelum}}
 \tablehead{
    \colhead{Phase \tablenotemark{b}} & 
    \colhead{\ion{O}{3} $\lambda$3760} &
    \colhead{H$\delta$/\ion{N}{3} $\lambda$4100} &
        \colhead{H$\gamma$} &
    \colhead{\ion{N}{3} $\lambda$4640} &
       \colhead{\ion{He}{2}} &
    \colhead{H$\beta$} & 
    \colhead{H$\alpha$} \\ 
  (days)     &  ($10^{41}$~erg~s$^{-1}$)      &     ($10^{41}$~erg~s$^{-1}$)      &   ($10^{41}$~erg~s$^{-1}$)     &   ($10^{41}$~erg~s$^{-1}$)       &   ($10^{41}$~erg~s$^{-1}$)       & ($10^{41}$~erg~s$^{-1}$) & ($10^{41}$~erg~s$^{-1}$)   }
\startdata
-24.2 &  1.16 ( 0.02) &  1.28 ( 0.02)&  0.49 ( 0.01) &  2.84 ( 0.02) &  0.51 ( 0.02) &  1.14 ( 0.03) &  1.23 ( 0.02) \\ 
 -7.5 &  1.13 ( 0.02) &  2.15 ( 0.01)&  1.34 ( 0.01) &  3.26 ( 0.03) &  0.79 ( 0.02) &  2.50 ( 0.01) &  2.56 ( 0.01) \\ 
  2.3 &  1.00 ( 0.03) &  2.06 ( 0.02)&  1.71 ( 0.01) &  2.75 ( 0.04) &  0.91 ( 0.03) &  2.66 ( 0.02) &  2.81 ( 0.02) \\ 
  7.3 &  0.90 ( 0.02) &  1.88 ( 0.01)&  1.44 ( 0.01) &  2.17 ( 0.03) &  0.96 ( 0.02) &  2.46 ( 0.01) &  3.00 ( 0.02) \\ 
 21.8 &    \nodata  &  1.35 ( 0.01)&  1.17 ( 0.01) &  1.50 ( 0.01) &  1.07 ( 0.01) &  2.10 ( 0.01) &  2.97 ( 0.01) \\ 
 34.7 &    \nodata  &  1.67 ( 0.01)&  1.34 ( 0.01) &  1.81 ( 0.02) &  1.41 ( 0.02) &  2.33 ( 0.02) &  4.02 ( 0.01) \\ 
 52.3 &     \nodata   &  0.41 ( 0.01)&  0.42 ( 0.01) &  0.61 ( 0.01) &  0.48 ( 0.01) &  0.76 ( 0.01) &  1.59 ( 0.01) \\ 
 67.1 &     \nodata   &  0.41 ( 0.01)&  0.33 ( 0.01) &  0.73 ( 0.01) &  0.32 ( 0.01) &  0.70 ( 0.01) &  1.27 ( 0.01) 
\enddata
\tablenotetext{a}{De-reddened only for Galactic extinction ($A_V = 0.625$~mag).}
\tablenotetext{b}{With respect to the date of $U$-band maximum ($\mathrm{MJD} = 58340.74$) and given in the rest frame of AT~2018dyb ($z=0.0180$).}
\end{deluxetable}

\begin{deluxetable}{rccccrr}
  \tablecaption{Emission line widths and velocity offsets \tablenotemark{a} \label{tab:FWHM}}
 \tablehead{
    \colhead{Phase \tablenotemark{b}} & 
    \colhead{\ion{O}{3} $\lambda$3760} &
    \colhead{H$\delta$/\ion{N}{3} $\lambda$4100} &
        \colhead{\ion{He}{2} }&
    \colhead{H$\alpha$}&
       \colhead{$v_{\lambda4100}$} &
    \colhead{$v_{H\alpha}$} \\ 
  (days)     &  ($10^{3}$~km~s$^{-1}$)      &     ($10^{3}$~km~s$^{-1}$)      &   ($10^{3}$~km~s$^{-1}$)     &   ($10^{3}$~km~s$^{-1}$)       &   (km~s$^{-1}$)       & (km~s$^{-1}$)  }
\startdata
-24.2 & 11.07 ( 0.15) & 12.36 ( 0.13)&  6.52 ( 0.19) & 13.31 ( 0.21) & -1061.61 (   49.56) &  -655.59 (   71.35) \\ 
 -7.5 & 12.09 ( 0.16) & 11.25 ( 0.05)&  9.45 ( 0.20) & 13.63 ( 0.05) &   -95.53 (   20.11) &   417.32 (   19.30) \\ 
  2.3 &  9.76 ( 0.20) &  9.09 ( 0.06)&  8.57 ( 0.23) & 12.71 ( 0.08) &   112.77 (   24.42) &   495.85 (   34.10) \\ 
  7.3 &  9.63 ( 0.17) &  9.20 ( 0.05)&  8.38 ( 0.17) & 12.54 ( 0.06) &   462.57 (   22.27) &   756.11 (   22.89) \\ 
 21.8 &   \nodata     &  6.38 ( 0.02)&  6.63 ( 0.05) & 10.96 ( 0.03) &   811.65 (   10.06) &   896.56 (   11.67) \\ 
 34.7 &   \nodata     &  5.81 ( 0.03)&  6.56 ( 0.07) &  9.58 ( 0.03) &   595.45 (   11.49) &   616.55 (    9.42) \\ 
 52.3 &   \nodata     &  4.92 ( 0.04)&  6.84 ( 0.10) &  7.86 ( 0.02) &   619.15 (   15.80) &   398.02 (    8.08) \\ 
 67.1 &   \nodata     &  6.33 ( 0.06)&  6.63 ( 0.13) &  7.33 ( 0.02) &   207.58 (   25.86) &   246.80 (    8.97) 
\enddata
\tablenotetext{a}{The FWHM of H$\gamma$ and H$\beta$ was constrained to be within 2,000~km~s$^{-1}$ of H$\alpha$ during the fit. Similarly,  \ion{N}{3} $\lambda$4640  was constrained by \ion{N}{3} $\lambda$4100. Therefore, no FWHMs are reported for these lines. Similarly, we only report the velocity offsets for the   \ion{N}{3} $\lambda$4100 and H$\alpha$ lines.}
\tablenotetext{b}{With respect to the date of $U$-band maximum ($\mathrm{MJD} = 58340.74$) and given in the rest frame of AT~2018dyb ($z=0.0180$).}
\end{deluxetable}

\end{document}